\documentclass[twocolumn,showpacs,noshowkeys,floatfix,prd,aps]{revtex4}
\usepackage{epsfig} 
\usepackage{longtable} 
\usepackage{hyperref}
\usepackage{bm}
\begin{document}  
\begin{titlepage} 
\null 
\vspace{2cm} 
\begin{center} 
\Large\bf  
ATLAS discovery potential for a heavy charged Higgs boson
\end{center} 
\vspace{1.5cm} 
 
\begin{center} 
\begin{large} 
K\'et\'evi~Adikl\`e~Assamagan\\ 
\end{large} 
\vspace{0.5cm} 
Department of Physics, Brookhaven National Laboratory\\ 
Upton, NY 11973 USA\\ 
\vspace{0.7cm} 
\begin{large} 
Yann Coadou\\ 
\end{large} 
\vspace{0.5cm} 
Department of Radiation Sciences,Uppsala University\\
Box 535, 751~21 Uppsala, Sweden\\
\vspace{0.7cm} 
\begin{large} 
Aldo Deandrea\\ 
\end{large} 
\vspace{0.5cm} 
Institut de Physique Nucl\'eaire, Universit\'e de Lyon I\\  
4 rue E.~Fermi, F-69622 Villeurbanne Cedex, France 
\end{center} 
 
\vspace{1.3cm} 
 
\begin{center} 
\begin{large} 
{\bf Abstract}\\[0.5cm] 
\end{large} 
\parbox{14cm}{The sensitivity of the ATLAS detector to the discovery
  of a heavy charged Higgs boson is presented. Assuming a heavy SUSY
  spectrum, the most promising channels above the top quark mass are
  $H^\pm\rightarrow tb$ and $H^\pm\rightarrow\tau^\pm\nu_\tau$ which
  provide coverage in the low and high $\tan\beta$ regions up to $\sim
  600$ GeV. The achievable precisions on the charged Higgs mass and
  $\tan\beta$ determination are also discussed. The $H^\pm\rightarrow
  W^\pm h^0$ channel, though restricted to a small MSSM parameter
  space, shows a viable signal in NMSSM where the parameter space is
  less constrained. The observation of the channel
  $H^-\rightarrow\tau^-_L\nu_\tau + c.c.$ may constitute a distinctive
  evidence for models with singlet neutrinos in large extra
  dimensions.}
\end{center}   
\vspace{2cm} 
\vfil 
\noindent 
LYCEN-2002-10\\
BNL-68986\\ 
ATL-COM-PHYS-2002-002\\
SN-ATLAS-2002-017\\
February 2002\\ 
\vfill 
\eject 
\end{titlepage} 
 
\newpage 
 
\title{ATLAS discovery potential for a heavy charged Higgs boson}  
\author{K\'et\'evi~A. Assamagan}  
\email{ketevi@bnl.gov} 
\affiliation{Department of Physics, Brookhaven National Laboratory, 
Upton, NY 11973 USA}
\author{Yann Coadou}
\email{yann@tsl.uu.se}
\affiliation{Department of Radiation Sciences,Uppsala University,
Box 535, 751~21 Uppsala, Sweden}
\author{Aldo Deandrea}
\email{deandrea@ipnl.in2p3.fr}
\affiliation{Institut de Physique Nucl\'eaire, Universit\'e Lyon I, 4 rue
E.~Fermi,  F-69622 Villeurbanne Cedex, France}   
\date{February, 2002}  
\preprint{LYCEN-2002-10}
\preprint{BNL-68986}  
\preprint{ATL-COM-PHYS-2002-002}  
\preprint{SN-ATLAS-2002-017}  

\def\OOrd{\lower .7ex\hbox{$\;\stackrel{\textstyle >}{\sim}\;$}}
\def\OOle{\lower .7ex\hbox{$\;\stackrel{\textstyle <}{\sim}\;$}}

\begin{abstract}    
The sensitivity of the ATLAS detector to the discovery of a heavy charged
Higgs boson is presented.  Assuming a heavy SUSY spectrum, the most promising
channels above the top quark mass are  $H^\pm\rightarrow tb$ and
$H^\pm\rightarrow\tau^\pm\nu_\tau$ which provide coverage  in the low and high
$\tan\beta$ regions up to $\sim 600$ GeV. The achievable precisions on the 
charged Higgs  mass and $\tan\beta$ determination are also discussed. The
$H^\pm\rightarrow W^\pm h^0$  channel, though restricted to a small MSSM
parameter space, shows a viable signal in NMSSM  where the parameter space is
less  constrained. The observation of the channel
$H^-\rightarrow\tau^-_L\nu_\tau + c.c.$ may constitute a distinctive evidence
for models with singlet neutrinos in large extra  dimensions. 
\end{abstract}  
 
\maketitle  

\section{Introduction}

In the Standard Model (SM), one scalar doublet is responsible for the
electroweak symmetry breaking, leading to the prediction of one
neutral scalar particle in the physical spectrum, the Higgs boson.
The spectrum of many extensions to the SM includes a charged Higgs
state.  We consider as a prototype of these models the two-Higgs
Doublet Model of type II (2HDM-II), where the Higgs doublet with
hypercharge $-1/2$ couples only to right--handed up--type quarks and
neutrinos whereas the $+1/2$ doublet couples only to right--handed
charged leptons and down--type quarks; an example is the Minimal
Supersymmetric Standard Model (MSSM). In the following we will use the
vacuum expectation values (VEV) $v\simeq 246$ GeV of the SM and $v_1$
(VEV of the $+1/2$ doublet) and $v_2$ (VEV of the $-1/2$ doublet) of
the 2HDM. They relate to each other as:
\begin{equation}
\frac{v}{\sqrt{2}}=\sqrt{v_1^2+v^2_2}~~~~~~~~ \tan \beta = \frac{v_2}{v_1}
\label{eq:vevs}
\end{equation}
and the tree level relation to the $W$ mass is $m_W^2=g^2 v^2/4=g^2 (v_1^2+v_2^2)/2$.
In the 2HDM models, the two complex Higgs doublets correspond to eight scalar
states. Symmetry breaking leads to  five Higgs bosons, three neutral (two
CP-even $h$, $H$ and one CP-odd $A$) and a  charged pair,
$H^\pm$~\cite{2HDM}. At tree level, the Higgs sector of the MSSM is specified
by two parameters, generally taken as $m_A$, the mass of the  CP-odd Higgs $A$
and $\tan\beta$, the ratio of the vacuum expectation values of the two
Higgs  doublets. However, radiative corrections can modify tree level
relations significantly --- the most  affected is the mass of the lightest
CP-even Higgs which is constrained at tree level to be below  $m_Z$ but with
radiative corrections, the upper bound is shifted to $\sim$
135~GeV~\cite{CONST}.  While the neutral Higgs bosons may be difficult to
distinguish from the one of the SM, the charged Higgs bosons are a distinctive
signal of physics beyond the SM. The detection of a $H^\pm$ may therefore play
an important role in the discovery of an extended Higgs sector, such as the one
required by the MSSM.

LEP searches have yielded a lower bound of 114.1~GeV
on the mass of the SM Higgs boson with a 2.1-$\sigma$ evidence for a 115.6~GeV
Higgs~\cite{LEP-SM}. An upper bound  of 196~GeV on the SM Higgs boson mass is
inferred from global fits to precision electroweak  data~\cite{GFIT}. At the
Tevatron Run 1, the SM Higgs boson has been searched for in the process 
$q\bar{q}\rightarrow Z(W)H$ where the associated vector boson provides a
suppression of the  backgrounds. These searches yield no evidence of the Higgs
as the observed events are consistent  with expectation from the
backgrounds~\cite{h_TeV}. The search for the SM Higgs boson will be continued
at the Tevatron where  the mass range covered will be
extended to $\sim$ 170~GeV with an integrated luminosity of up to 
$\sim$~40~fb$^{-1}$. At the LHC (Large Hadron Collider), a SM Higgs signal can be observed 
with a significance of  more than 5$\sigma$ after just several months of data taking
($<$ 10~fb$^{-1}$). 

In the MSSM, lower bounds of 91.0 and 91.9~GeV on the masses of the lightest
CP-even Higgs $h$  and the CP-odd Higgs $A$ have been set in the experimental
searches at LEP. Further, the  $\tan\beta$ regions of $0.5 < \tan\beta < 2.4$
and $0.7 < \tan\beta < 10.5$ are excluded at 95\%  confidence level for the
maximum $m_h$ and the no-mixing scenarios respectively~\cite{LEP-SM}. In addition, 
a large area of the 2HDM-II parameter space has been scanned leading to the exclusion 
at 95\% CL of significant regions of the ($m_h$, $m_A$), ($m_h$, $\tan\beta$) and 
($m_A$, $\tan\beta$) projections. Within the scanned parameter space, the region 
$1 \OOle m_h \OOle 44$~GeV and $12 \OOle m_A \OOle 56$~GeV is excluded at 95\% CL 
independent of $\tan\beta$ and $\alpha$~\cite{OPAL}. A model-independent interpretation, 
with no assumption made on the structure of the Higgs sector, was also conducted at LEP. 
Lower bounds are set at 95\% CL on the masses of  the scalar and pseudo-scalar neutral 
Higgs bosons $S^0$ and $P^0$ --- in the search for the generic processes $e^+e^- \to S^0Z^0$ 
and $e^+e^- \to S^0P^0$ --- depending on the assumed values of the scale factors $s^2$ and 
$c^2$, and of the branching ratios~\cite{OPAL}. At the Tevatron, the lightest Higgs boson 
of MSSM has been searched for in the process
$p\bar{p}\rightarrow  b\bar{b}h\,\, (h\rightarrow b\bar{b})$. These searches
excluded the large $\tan\beta$ region  ($\tan\beta > 35$) not accessible at
LEP~\cite{h_TeV}.

In the MSSM, the charged Higgs mass at tree level, $m_{H^\pm}$, is related to 
$m_A$ as:
\begin{equation}
\label{eq:mA_mHp}
m_{H^\pm}^2 = m_W^2 + m_A^2 \; ,
\end{equation}
and is less sensitive to radiative corrections~\cite{Hp_RAD}. At LEP, the
main production mechanism of the charged Higgs is $e^+e^-\rightarrow H^+H^-$. Direct searches
of the charged Higgs at  LEP have been carried in the general 2HDM (where $m_{H^\pm}$ is
not constrained) assuming $H^+\rightarrow\tau^+\nu_\tau\, (c\bar{s})$  and
$H^-\rightarrow\tau^-\bar{\nu}_\tau\, (\bar{c}s)$. These searches yielded a lower bound of
78.6~GeV on the  charged Higgs mass independent of the
$H^\pm\rightarrow\tau^\pm\nu_\tau$ branching  ratio \cite{hp_LEP}. At the
Tevatron, CDF and D{\O} performed direct and indirect searches for the 
charged Higgs through the process $p\bar{p}\rightarrow t\bar{t}$, with at
least one top quark  decaying via $t\rightarrow H^\pm b$. The direct searches
seek the process  $H^\pm\rightarrow\tau^\pm\nu_\tau$ with the identification of
the $\tau$ lepton through its hadronic  decays. In the indirect searches, CDF
and D{\O} looked for a deficit in the SM $t\bar{t}$ decays  caused by the
possible existence of $t\rightarrow H^\pm b$. These searches excluded the low
and  high $\tan\beta$ regions up to $\sim$ 160~GeV~\cite{hp_TEV}. Other
experimental bounds on the  charged Higgs mass come from processes where the
charged Higgs enters as a virtual particle. One  such process is the
$b\rightarrow s\gamma$ decay where indirect limits are obtained from the 
measurement of the decay rate~\cite{CLEO}. However, these bounds are strongly
model  dependent~\cite{Gambino:2001ew,BORZ,COAR}. 

\begin{figure}  
\epsfysize=5truecm
 \begin{center}  
\epsffile{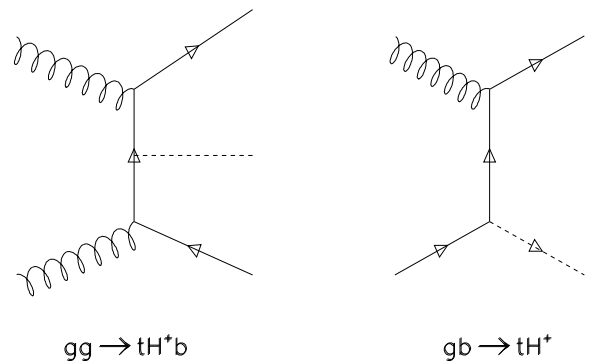}  
\caption{The charged Higgs production
at the LHC through the $2\rightarrow 3$ process,   $gg\rightarrow tbH^\pm$ and
the $2\rightarrow 2$ process, $gb\rightarrow  tH^\pm$. The inclusive 
cross section is the sum of both contributions after  the subtraction of the common
terms.}  \label{fig:prod_graphs}  
\end{center}  
\end{figure}

\begin{figure} 
\epsfysize=8truecm  
\begin{center}  
\epsffile{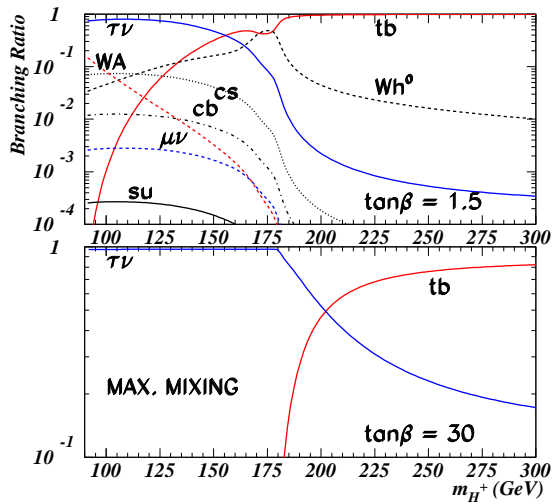}  
\caption{The branching ratios of the charged decays in SM particles as a
function of $m_{H^\pm}$ for  $\tan\beta=1.5$ (top plot) and $\tan\beta=30$ (bottom plot). 
The most dominant decay channels are $H^\pm\rightarrow\tau^\pm\nu_\tau$ and  
$H^\pm\rightarrow tb$.} 
\label{fig:HDECAYS}   
\end{center}  
\end{figure}

The search for the charged Higgs boson
will be continued above the top quark mass. The main  production mechanisms
would be the $2\rightarrow 3$ process $gg\rightarrow tbH^\pm$ and the $2\rightarrow 2$ 
process $gb\rightarrow tH^\pm$ shown in Fig.~\ref{fig:prod_graphs}~\cite{Hp-PROD}.  
Additional production mechanisms come from the Drell-Yan type process $gg,\,
q\bar{q}\rightarrow  H^+H^-$~\cite{Hp-PROD1}, and the associated production
with a $W$ boson, $q\bar{q}\rightarrow  H^\pm W^\mp$~\cite{Hp-PROD2}. However,
in the former case, the rate is rather low at the LHC either because of weak
couplings and low quark luminosity or the process is induced by loops of 
heavy quarks and therefore suppressed by additional factors of electroweak
couplings; in the latter case, the rate is also somewhat lower at the LHC and this channel 
suffers from the large irreducible $t\bar{t}$ and QCD jet backgrounds~\cite{WH}. The main  
production mechanisms, i.e., the $2\rightarrow 3$ and the
$2\rightarrow 2$ processes, partially overlap  when the former is obtained from
the latter by a gluon splitting into a $b$-quark pair. When summing  both
contributions, care must be taken to avoid double counting. The difference
between the two  processes is well understood and the inclusive cross section
is obtained from a proper subtraction of  the common logarithmic
terms~\cite{subtraction,3b,reviewLHC,Jaume}. Assuming a heavy SUSY  spectrum,
the charged Higgs will decay only into SM particles as shown in 
Fig.~\ref{fig:HDECAYS}~\cite{HDECAY} for the maximal stop mixing scenario. For low
values of  $\tan\beta$, below the top quark mass, the main decay channels are 
$H^\pm\rightarrow\tau^\pm\nu_\tau$, $c\bar{s}$, $Wh^0$ and $t^*b$; above the
top quark mass,  the $H^\pm\rightarrow tb$ decay mode becomes dominant. For
high values of $\tan\beta$, the  $H^\pm\rightarrow\tau^\pm\nu_\tau$ and
$H^\pm\rightarrow tb$ are the only dominant decay modes.   

In this paper, we summarise the sensitivity of the ATLAS detector at the LHC to
the discovery of a  heavy charged Higgs, with emphasis on the region above the 
top quark mass, $m_{H^\pm} > m_t$ (some of the results presented here have already been 
published in earlier papers~\cite{ASSA1,ASSA2,ASSA3,ASSA4,ASSA5}).

These studies have been carried out as particle level  event generation in PYTHIA5.7 and
PYTHIA6.1~\cite{PYTHIA}, at $\sqrt{s}=14$~TeV, with the  detector resolutions
and efficiencies parameterised in ATLFAST~\cite{ATLFAST} from full 
detector simulations. We used the CTEQ2L and CTEQ5L parton distribution function 
parametrisations~\cite{CTEQ} and the charged Higgs mass is calculated to 1-loop with
FeynHiggsFast~\cite{FeynHiggs}. 

In section \ref{sec:det}, we describe briefly the ATLAS detector and the performance of the 
detector components necessary for the discovery of the charged Higgs discovery. In 
section~\ref{sec:tb}, we discuss the  possibility to detect the process 
$H^\pm\rightarrow tb$, followed by $H^\pm\rightarrow\tau^\pm\nu_\tau$ in section 
\ref{sec:taunu}. Then, we discuss the process $H^\pm\rightarrow W^\pm h^0$ and we give the 
expected achievable  precisions on the charged Higgs mass and $\tan\beta$
determination. The detection of a charged Higgs  signal in models with singlet
neutrinos in large extra dimensions is discussed in section \ref{sec:extra}.
Then, a  discussion on charged Higgs decay to supersymmetric particles and
charged Higgs production from  SUSY cascade decays is presented followed by
concluding remarks. 

\section{Detector description and performance}
\label{sec:det}

The ATLAS detector is a general purpose detector designed and optimised to be sensitive 
to a wide range of physics issues to be explored at the LHC such as the origin of mass at 
the electroweak scale. The detector itself consists of an inner detector, electromagnetic 
and hadronic calorimeters, a stand-alone muon spectrometer and a magnet system. 

The inner detector comprises discrete high resolution semiconductor pixel and strip 
detectors in the inner section, and a straw tube tracking detector with the capability 
for transition radiation detection in the outer part. The inner detector provides pattern 
recognition, momentum and vertex measurements, and electron identification.  

The calorimeter consists of a highly segmented electromagnetic (EM) sampling calorimeter 
followed by a hadronic calorimeter (HAD). The EM calorimeter is a lead/liquid argon 
detector with accordion shaped kapton electrodes and lead absorber plates providing 
electron and photon identification and measurements. This is complemented by hermetic 
hadronic calorimeters for jets and missing energy  measurements. The HAD covers the range 
$|\eta|< 5$ with different detector technologies best suited to the variety of 
requirements and radiation environments. They consist of a barrel sampling detector using 
iron as the absorber and scintillating tiles as the active material, end-cap calorimeters 
with copper/liquid argon technology, and a forward calorimeter using liquid argon with 
rod-shaped electrodes in a tungsten matrix. 

The calorimeter is surrounded by a stand-alone muon spectrometer whose design is based on 
the deflection of muon tracks in large super-conducting magnets. It consists of four 
different chamber technologies, two of which are for precision measurements of muon track 
parameters and the other two for triggering. In the barrel section, resistive plate 
chambers provide the trigger function whereas in the end-caps, this is done by thin gap 
chambers. Precision measurements of tracks are done with monitored drift tubes, however 
at large pseudo-rapidity and close to the interaction point, highly granular cathode 
strip chambers are used to withstand the rate and the background conditions. The muon 
spectrometer provides precision measurements of muon momenta and the capability for 
low-$p_T$ trigger at low luminosity.

The magnet system consists of a central solenoid which provides a 2~Tesla magnetic field 
for the inner detector, surrounded by a system of three large super-conducting air-core 
toroids generating the magnetic field for the muon spectrometer.  

Further details on the detector design and optimisation including the trigger and data 
acquisition system are well documented in~\cite{TP,TDR}. The detection of a charged Higgs 
signal would depend on many crucial ATLAS detector performance parameters, 
namely~\cite{TDR,BOSM}: 
\begin{itemize}
\item $\tau$-jet reconstruction and rejection against QCD jets (for a $\tau$-jet reconstruction 
efficiency of 30\%, a jet rejection factor of $\sim$ 400 can be achieved).
\item Good $E_T^\mathrm{ miss}$ resolution as the $p_T^\mathrm{ miss}$ vector and the reconstructed 
$\tau$-jet will be used for the transverse mass reconstruction of 
$H^\pm\rightarrow\tau^\pm\nu_\tau$ --- $E_T^\mathrm{ miss}$ resolutions of 20-100~GeV are expected 
based on full detector simulations of $A/H\rightarrow\tau\tau$ events in the mass range 
100-500~GeV. 
\item A $b$-tagging performance of 60\% (50\%) at low (high) luminosity is expected and necessary 
for the reconstruction $H^\pm\rightarrow tb$ and $H^\pm\rightarrow W^\pm h^0\, (\to b\bar{b})$ 
which contain several $b$-jets in the final state. 
\item Excellent jet reconstruction and calibration would also be needed as the reconstruction 
of $W^\pm\to jj$ is necessary for the observation of the signals studied herein.
\item Finally an isolated lepton ($e$ or $\mu$) trigger is needed for $H^\pm\to tb$ and 
$H^\pm\to W^\pm h^0$ and a lepton identification efficiency of 90\% is expected. For the 
$H^\pm\to\tau^\pm\nu_\tau$, a multi-jet trigger (and possibly a $\tau$ trigger) would be necessary. 
Such a trigger will be available, not just for the charged Higgs, but also for many other 
important physics studies.
\end{itemize}

The ATLFAST~\cite{ATLFAST} simulation code used for the present work has been carefully 
verified with fully reconstructed results. Performance figures are well reproduced and we
expect only slightly worse performance with the real ATLAS detector.

\section{The $\bm{H^\pm\rightarrow tb}$ channel}
\label{sec:tb}

The region $m_{H^\pm} > m_t$ was at first considered problematic, as the large decay mode
$H^\pm \to tb$ has large QCD backgrounds at hadron colliders. However the possibility of 
efficient $b$-tagging has considerably improved the situation \cite{barger}. The 
interaction 
term of the charged Higgs with the $t$ and $b$ quarks in the 2HDM of type II is
given by:
\begin{eqnarray}
{\cal L} &= &\frac{g}{2\sqrt{2}\; m_W} V_{tb} H^+ {\bar t} \left( m_t \cot \beta
(1-\gamma_5)  \right.\nonumber \\
&+&\left. m_b \tan \beta (1+\gamma_5) \right) b + h.c.
\end{eqnarray}
We consider therefore the large $2\rightarrow 2$ hadroproduction process
$gb\rightarrow t H^\pm$  (see Fig.~\ref{fig:prod_graphs}) with decay
mechanism $H^\pm\rightarrow tb$. The cross section for $gb\rightarrow tH^\pm$ can be 
written as:
\begin{equation}
\label{gbcross}
\sigma(gb\rightarrow tH^\pm) \propto m^2_t\cot^2\beta + m^2_b\tan^2\beta.
\end{equation}
The decay width of $H^- \to {\bar t}b$ is given by:
\begin{eqnarray}
&&\Gamma(H^- \to {\bar t}b) \simeq \frac{3 \; m_{H^\pm}}{8 \; \pi v^2} 
\Big[ \left( m_t^2 \cot^2 \beta + m_b^2 \tan^2 \beta \right) \nonumber \\
&\times& \left( 1 -\frac{m_t^2}{m_{H^\pm}^2}-\frac{m_b^2}{m_{H^\pm}^2}\right) - 
\frac{4 m_t^2 m_b^2}{m_{H^\pm}^2} \Big] \nonumber \\
&\times& \left[ 1 -\left(\frac{m_t+m_b}{m_{H^\pm}}\right)^2 \right]^{1/2}
\left[ 1 -\left(\frac{m_t-m_b}{m_{H^\pm}}\right)^2 \right]^{1/2}
\label{decayhtb}
\end{eqnarray}
where the factor 3 takes into account the number of colours.
The final state of the $2\rightarrow 2$ hadroproduction process contains two top quarks, 
one of which is required to decay semi-leptonically to provide the trigger, 
$t\rightarrow l\nu b$ ($l=e,\,\,\mu$) and the other hadronically,
$\bar{t}\rightarrow jjb$. The main  background comes from $t\bar{t}b$ and
$t\bar{t}q$ production with $t\bar{t}\rightarrow  WbWb\rightarrow l\nu b jj
b$. The rates for the signal and the backgrounds are shown in Table~\ref{tab:tb_1}. 
\begin{table*}
\begin{center}
\begin{minipage}{.75\linewidth} 
\caption{\label{tab:tb_1} The expected rates ($\sigma\,\times\,$~BR, in pb) --- for the signal 
$bg\rightarrow H^\pm t\rightarrow l\nu bjjbb$ and the $t\bar{t}b\,+t\,\bar{t}q$ 
background for several values of $m_{H^\pm}$ and $\tan\beta$.}
\end{minipage}
\vbox{\offinterlineskip 
\halign{&#& \strut\quad#\hfil\quad\cr  
\colrule
& Process   && $m_{H^\pm}$ (GeV)  && $\tan\beta=1.5$   && $\tan\beta=10$ && $
\tan\beta=30$ & 
\cr
\colrule
& $bg\rightarrow H^\pm t\rightarrow l\nu bjjbb$  && $200$   &&   3.4  &&   0.4 
&& 1.6  & \cr
&       && $250$     &&   2.0       &&   0.18    && 1.2           & \cr
&       && $300$     &&   1.2       &&   0.14    && 1.0           & \cr
&       && $400$     &&   0.54      &&   0.08    && 0.4          & \cr
&       && $500$     &&   0.3       &&   0.04    && 0.2           & \cr
\colrule
& $t\bar{t}\rightarrow jjbWb\, (W\rightarrow l\nu)$  &&   &&   228     &&   228    
&& 228  & \cr
\colrule}}
\end{center}
\end{table*}

We search for an isolated lepton, three $b$-tagged jets and at least two non $b$-jets. We 
retain the jet-jet combinations whose invariant masses are consistent with the $W$-boson mass, 
$|m_W-m_{jj}| < 25$~GeV (for the events satisfying this requirement, the 4-momenta of the jets 
are rescaled such that $m_{jj}=m_W$) and we use the $W$-boson mass constraint to find the 
longitudinal component of the neutrino momentum, $W^\pm\to l\nu$, assuming the missing transverse 
momentum is the neutrino transverse momentum. Subsequently, the two top quarks in the spectrum 
are reconstructed, retaining the pair of top quarks whose invariant masses $m_{l\nu b}$ and 
$m_{jjb}$ best minimise $\chi^2 \equiv (m_t-m_{l\nu b})^2 + (m_t - m_{jjb})^2$. 
The remaining $b$-jet can be paired with either top quark to give two charged Higgs candidates, 
one of which leads to a combinatorial background. 

\begin{figure}
\epsfxsize=8truecm
\begin{center}
\epsffile{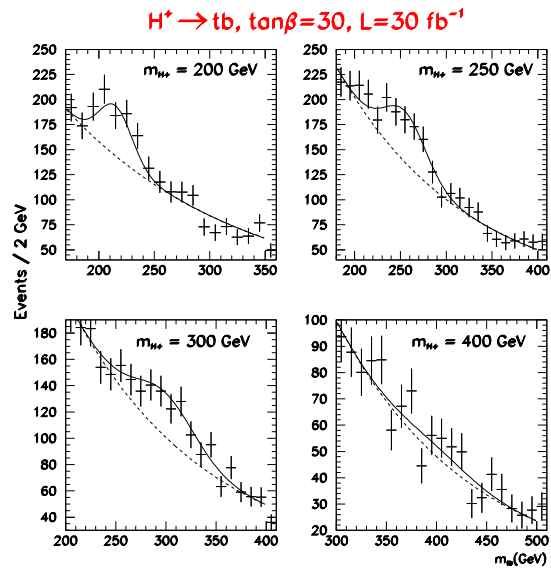}
\caption{The signal$+$background (solid line) and the background (dashed line) distributions 
for the reconstructed invariant mass $m_{tb}$ of a Higgs mass of 200, 250, 300 and 400~GeV, 
$\tan\beta=30$ and an integrated luminosity of 30~fb$^{-1}$. The errors are statistical only.}
\label{fig:tb_sig6}
\end{center}
\end{figure}

Above $m_{H^\pm}=300$~GeV, the reduced signal 
rate and the combinatorial background make the observation of this channel difficult. Below a 
charged Higgs mass of 300~GeV, this channel may be observed above the $t\bar{t}b$ plus 
$t\bar{t}q$ background. The results for $m_{H^\pm}=200-400$~GeV are shown in 
Fig.~\ref{fig:tb_sig6}.
\begin{table*}
\begin{center}
\begin{minipage}{.75\linewidth} 
\caption{\label{tab:tb_5} Sensitivity of the ATLAS detector ($S/\sqrt{B}$) to the observation of 
the charged Higgs through $H^\pm\rightarrow tb$. Discovery is possible in the low ($<2.5$) and 
the high ($>25$) $\tan\beta$ regions up to 400~GeV.}
\end{minipage}
\vbox{\offinterlineskip 
\halign{&#& \strut\quad#\hfil\quad\cr  
\colrule
& $m_{H^\pm}$ (GeV) && $\tan\beta=1$ && $\tan\beta=2$ &&  $\tan\beta=10$ && $\tan\beta=25$ 
&& $\tan\beta=35$ & \cr
\colrule
& 200          && 11.5     && 5.3      &&  1.3    &&  2.9     && 5.5        & \cr
& 250          && 19.6     && 6.1      &&  1.1    &&  5.1     && 11.1       & \cr
& 300          && 13.8     && 5.2      &&  1.1    &&  4.9     && 9.9        & \cr
& 400          && 7.7      && 2.8      &&  0.5    &&  2.3     && 4.7        & \cr
\colrule}}
\end{center}
\end{table*}
At high values of $\tan\beta$ ($>$ 25), sensitivity is expected up to 400~GeV as 
shown in Table~\ref{tab:tb_5}. The 5-$\sigma$ discovery contour for $H^\pm\rightarrow tb$ is 
shown in Fig~\ref{fig:contour}. This analysis is presented extensively in~\cite{ASSA1}.

\section{The $\bm{H^\pm\rightarrow\tau^\pm\nu_\tau}$ channel}
\label{sec:taunu}
The $\tau \nu$ decay channel offers a high $p_T$ of the $\tau$ and large
missing energy signature that can be discovered at LHC over a vast region of
the parameter space \cite{oda}. The events are generated in PYTHIA using the
process $gb\rightarrow t H^\pm$. The associated  top quark is required to decay
hadronically, $t\rightarrow jjb$. The charged Higgs decays into a  $\tau$
lepton, $H^\pm\rightarrow\tau^\pm\nu_\tau$, and the hadronic decays of the
$\tau$ are  considered. The backgrounds considered are QCD, $W+$jets, single
top production  $Wt$, and $t\bar{t}$, with one $W\rightarrow jj$ and the other
$W^\pm\rightarrow\tau^\pm\nu_\tau$. Table~\ref{tab:taunu_1} shows the rates
for the signal and the backgrounds as a function of $m_{H^\pm}$ and
$\tan\beta$.  \begin{table*} \begin{center} \begin{minipage}{.6\linewidth} 
\caption{\label{tab:taunu_1} The expected rates ($\sigma\times BR$), for the
signal  $gb\rightarrow t H^\pm$ with \mbox{$H^\pm\rightarrow\tau^\pm\nu_\tau$}
and $t\rightarrow jjb$,  and for the backgrounds: QCD, $W+$jets, $Wtb$ and
$t\bar{t}$ with $t\rightarrow\tau\nu b$ and  $\bar{t}\rightarrow jjb$. We
assume an inclusive $t\bar{t}$ production cross section of 590~pb.  Other
cross sections are taken from PYTHIA. The branching fractions of 
$H^\pm\rightarrow\tau^\pm\nu_\tau$  are obtained from HDECAY~\cite{HDECAY},
and we take the $W\rightarrow jj$ branching ratio to be  $2/3$.} \end{minipage}
\vbox{\offinterlineskip 
\halign{&#& \strut\quad#\hfil\quad\cr  
\colrule
& Process                && $\tan\beta$ && $m_{H^\pm}$~(GeV) && $\sigma\,\times\,$~BR (pb) & \cr
\colrule
& Signal                 &&   15        &&   180             &&    1.33    & \cr
&                        &&   30        &&   200             &&    2.23    & \cr
&                        &&   40        &&   250             &&    0.91    & \cr
&                        &&   45        &&   300             &&    0.54    & \cr
&                        &&   25        &&   350             &&    0.10    & \cr
&                        &&   35        &&   400             &&    0.13    & \cr
&                        &&   60        &&   450             &&    0.23    & \cr
&                        &&   50        &&   500             &&    0.11    & \cr
& $t\bar{t}$             &&             &&                   &&    84.11   & \cr
& $Wt$ ($p_T>30$~GeV) &&        &&                   &&    56.9    & \cr
& $W+$jets ($p_T>30$~GeV) &&            &&                   &&    $1.64\,10^4$ & \cr
& QCD ($p_T>10$~GeV)     &&             &&                   &&    $6.04\,10^9$ & \cr
 \colrule}}
\end{center}
\end{table*}

The width of the process $H^\pm\rightarrow\tau^\pm\nu_\tau$ is:
\begin{eqnarray}
&&\Gamma(H^- \to \tau^- \nu_\tau) \simeq \frac{m_{H^\pm}}{8 \; \pi v^2} 
\Big[ m_\tau^2 \tan^2 \beta  \nonumber \\
&\times& \left( 1 -\frac{m_\tau^2}{m_{H^\pm}^2}\right)  \Big]
\left( 1 -\frac{m_\tau^2}{m_{H^\pm}^2} \right) \; .
\label{decayhtaunu}
\end{eqnarray}
If the decay $H^\pm \to tb$ is kinematically allowed, comparing (\ref{decayhtaunu}) with 
(\ref{decayhtb}) one can have a rough estimate of the $H^\pm\rightarrow\tau^\pm\nu_\tau$ 
branching ratio $BR_\tau$:
\begin{eqnarray}
\label{eq:BR}
BR_\tau & \simeq & \frac{\Gamma(H^\pm\rightarrow\tau^\pm\nu_\tau)}{\Gamma(H^\pm\rightarrow 
tb)+
\Gamma(H^\pm\rightarrow\tau^\pm\nu_\tau)} \nonumber \\
 & = & \frac{m^2_\tau\tan^2\beta}{3(m^2_t\cot^2\beta+m^2_b\tan^2\beta) + m^2_\tau\tan^2\beta}.
\end{eqnarray}
A measurement of the signal rate in $H^\pm \to \tau^\pm\nu_\tau$  can allow a determination of 
$\tan \beta$ (see section \ref{detbeta} for details).

The distributions of one-prong hadronic decays of $\tau$'s, 
\begin{eqnarray}  
\label{eq:pinu} 
\tau^\pm  \rightarrow & \pi^\pm\nu_\tau  & ~~~(11.1\%) \\
\tau^\pm  \rightarrow & \rho^\pm(\rightarrow\pi^\pm\pi^0)\nu_\tau  & ~~~(25.2\%) \nonumber \\
\tau^\pm  \rightarrow & a_1^\pm(\rightarrow\pi^\pm\pi^0\pi^0)\nu_\tau & ~~~(9.0\%), \nonumber  
\end{eqnarray} 
are sensitive to the polarisation state of the $\tau$ lepton~\cite{POL1,POL2,POL3}. In fact, 
it is 
to be noted that the spin state of $\tau$'s coming from $H^\pm$- and $W^\pm$-boson decays are 
opposite as illustrated in Fig.~\ref{fig:tau_pol_led}. This is true for the case of one-prong 
decays into both $\pi^\pm$'s and longitudinal vector mesons, while the transverse component of the 
latter dilutes the effect and must be somehow eliminated by requiring that 80\% of the $\tau$-jet 
(transverse) energy is carried away by the $\pi^\pm$'s, i.e.: 
\begin{equation}
\label{eq:pfrac}
R=\frac{p^{\pi^\pm}}{p_T^{\tau}}> 0.8.
\end{equation}  
Alternatively, one can demand a hard distribution in $\Delta p_T$ which is the difference in the 
momenta of the charged track and the accompanying neutral pion(s)~\cite{POL1},
\begin{equation}
\Delta p_T = |p_T^{\pi^\pm}-p_T^{\pi^0}|.
\end{equation}
Ultimately, the polarisation effect leads to a significantly harder momentum distribution of 
charged pions from $\tau$ decays for the $H^\pm$ signal compared to the $W^\pm$ background, which 
can then be exploited to increase the signal-to-background ratios and the signal 
significances~\cite{POL1,POL2,POL3} by reducing the background and enhancing the signal~\cite{ASSA2}. We have included the $\tau$ polarisation into PYTHIA 
through the TAUOLA~\cite{TAUOLA} simulation code and considered all the hadronic decays of the 
$\tau$ lepton.

We search for one hadronic $\tau$ jet and at least three non $\tau$ jets, one of which must be a 
$b$-tagged jet. Further, we apply a $b$-jet veto to reject the $t\bar{t}$ background. As there is 
no isolated lepton (electron or muon) in the final state, the observation of this channel 
requires a multi-jet trigger with a $\tau$ trigger.
\begin{figure}
\epsfxsize=8truecm
\begin{center}
\epsffile{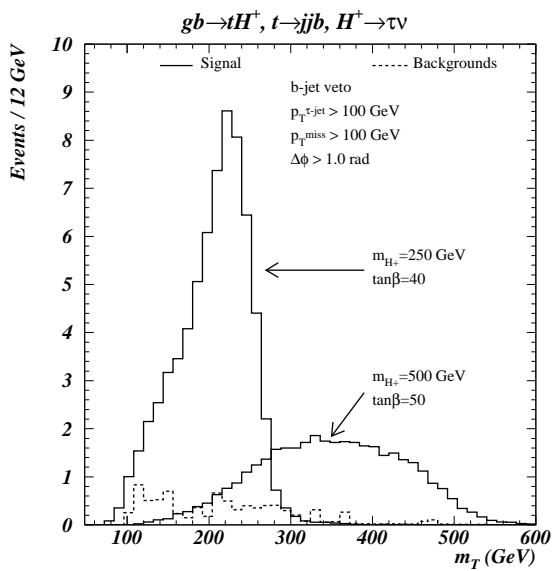}
\caption{The final transverse mass $m_T$ reconstruction for the signal and the backgrounds taking 
into account the polarisation of the $\tau$ lepton, for an integrated luminosity of 30~fb$^{-1}$.}
\label{fig:taunu5}
\end{center}
\end{figure}
After reconstructing the jet-jet invariant mass $m_{jj}$ and retaining the candidates consistent 
with the $W$-boson mass, the jet 4-momenta are rescaled as done in the $H^\pm\to tb$ analysis and 
the associated top quark is reconstructed by minimising $\chi^2 \equiv (m_{jjb}-m_t)^2$. 
Subsequently, a sufficiently high threshold on the $p_T$ of the $\tau$ jet is required. The 
background events satisfying this cut need a large boost from the $W$ boson. This results in a 
small azimuthal opening angle $\Delta\phi$ between the $\tau$ jet and the missing momentum 
${p\!\!\!/}_T$. In contrast, such a boost is not required from the $H^\pm$ for the signal 
events, leading to a backward peak in the azimuthal opening angle~\cite{ASSA2}. Furthermore, the missing 
momentum is harder for the signal. The difference between signal and background distributions 
in the azimuthal angle and the missing momentum increases with increasing $m_{H^\pm}$. These 
effects are well cumulated in the transverse mass 
\begin{equation}
\label{eq:trans}
m_T = \sqrt{2p_T^\tau{p\!\!\!/}_T\left[1-\cos(\Delta\phi)\right]},
\end{equation} 
which provides a good discrimination between the signal and the 
backgrounds as shown in Fig~\ref{fig:taunu5} --- the full invariant mass is not reconstructed in 
this case because of the neutrino in the final state. For the backgrounds, the transverse mass 
is kinematically constrained to be smaller than $m_W$ but for 
the signal, the transverse mass is bound from above by $m_{H^\pm}$. However, because of the 
experimental resolution of $E_T^\mathrm{ miss}$, some leakage of the background events 
into the signal region is observed (see Fig.~\ref{fig:taunu5}). The backgrounds in this channel 
are relatively small; 
significances upwards of $5\sigma$ can be achieved for $m_{H^\pm} > m_t$ and $\tan\beta > 10$, 
for an integrated luminosity of 30~fb$^{-1}$ as shown in Table~\ref{tab:taunu_3}. 
\begin{table*}
\begin{center}
\begin{minipage}{.7\linewidth} 
\caption{\label{tab:taunu_3} The expected signal-to-background ratios and significances 
calculated after all cuts for an integrated luminosity of 30~fb$^{-1}$. The backgrounds are 
relatively small; in fact, it is the size of the signal itself that limits the range of the 
discovery potential.}
\end{minipage}
\vbox{\offinterlineskip 
\halign{&#& \strut\quad#\hfil\quad\cr  
\colrule
& $\tan\beta$       & 30    & 40    & 45     & 25     & 35     & 60     & 50 & \cr
& $m_{H^\pm}$ (GeV) & 200   & 250   & 300    & 350    & 400    & 450    & 500 & \cr
\colrule
& Signal events     & 46.3  & 60.3  & 70.5   & 18.8   & 30.6   & 66.9   & 36.2 & \cr 
\colrule
&$t\bar{t}$          & 3.1  & 3.1  & 3.1   & 3.1   & 3.1   & 3.1   & 3.1 & \cr
& $Wt$       & 3.2  & 3.2  & 3.2   & 3.2   & 3.2   & 3.2   &  3.2 & \cr
& $W+$jets          & 0.3  & 0.3  & 0.3   & 0.3   & 0.3   & 0.3   &  0.3 & \cr 
\colrule
& Total background  & 6.7  & 6.7  &  6.7  & 6.7   &  6.7  & 6.7   &  6.7 & \cr 
\colrule
& $S/B$             & 6.9  & 9.0  & 10.5   & 2.8   &  4.6  & 10.0   &  5.4  & \cr
& $S/\sqrt{B}$      & 17.9  & 23.3  & 27.2   & 7.3   &  11.8  & 25.8   &  14.0  & \cr
\colrule}}
\end{center}
\end{table*}
The discovery contour for this channel is shown 
in Fig.~\ref{fig:contour}. In fact, the range of discovery potential is solely limited by the 
signal size itself. The present study shows a statistically significant improvement in the 
signal-to-background ratios and in the signal significances due to the $\tau$ polarisation 
effect but it is not necessary to restrict oneself to just the one-prong decays. Details of this study are available in~\cite{ASSA2}.

\section{The $\bm{H^\pm\rightarrow W^\pm h^0}$ channel}
\label{sec:wh}

Thus far, the study of the discovery potential of the charged Higgs with ATLAS has concentrated 
mainly on the fermionic decays modes --- $H^\pm\rightarrow tb$ and 
$H^\pm\rightarrow\tau^\pm\nu_\tau$~\cite{ASSA1,ASSA2,MOR-WH01}. In this section, the 
discovery potential of the charged Higgs with the ATLAS detector through the process 
$H^\pm\rightarrow W^\pm h^0$ is studied. The decay width of $H^- \to W^- h^0$ is:
\begin{eqnarray}
&&\Gamma(H^- \to W^- h^0) \simeq \frac{\cos^2 (\beta -\alpha)}{16 \; \pi v^2} 
\nonumber \\
&\times& \left[ 1 -\left(\frac{m_h+m_W}{m_{H^\pm}}\right)^2 \right]^{1/2}
\left[ 1 -\left(\frac{m_h-m_W}{m_{H^\pm}}\right)^2 \right]^{1/2}
\label{decayhwh}
\end{eqnarray}
where $\alpha$ is the Higgs mixing angle in the CP-even sector. In the MSSM one can easily verify that the 
maximum value of $\cos^2 (\beta -\alpha)$ at fixed $\tan \beta$ is $\cos^2 2\beta$
and occurs in the limit $m_{H^+} \to m_W$. Therefore the expectation of a heavy 
charged Higgs implies a small coupling for this channel. Though significant only in a tiny range 
of MSSM parameter space~\cite{MOR-WH02}, this channel constitutes a unique test for the MSSM and 
is also sensitive to the next-to-minimal extension to the MSSM, i.e., NMSSM, where there may be a 
significant range of viability  below and above the top quark mass~\cite{MOR-WH03}. 

\subsection{$\bm{H^\pm\rightarrow W^*h^0}$, $\bm{m_{H^\pm} < m_t}$}

Below the top quark mass, we consider $t\bar{t}$ production with one top quark decaying into a 
$W$ boson and the other into the charged Higgs. The characteristics of the production and decay 
processes are:
\begin{eqnarray}
gg\,(q\bar{q}) & \rightarrow & t\bar{t} \nonumber \\
t              & \rightarrow & H^\pm\,b \nonumber \\
\bar{t}        & \rightarrow & W\,\bar{b} \nonumber \\
H^\pm          & \rightarrow & W^*\,h^0. 
\end{eqnarray} 
Thus, the spectrum contains two $W$ bosons, one of which is off mass shell, and four $b$-quarks 
due to the subsequent decay $h^0\rightarrow b\bar{b}$. In the present analysis, one of the $W$'s 
is required to decay into leptons ($e$, $\mu$), and the other into jets. The major background to 
this process comes from $t\bar{t}b\bar{b}$ and $t\bar{t}q\bar{q}$ production where both top 
quarks decay into $W$'s. We search for an isolated lepton, four $b$-tagged jets and at least two 
non $b$-jets. Two possible scenarios are considered on an event by event basis:
\begin{eqnarray}
W^*\rightarrow l\nu &           & W\rightarrow jj \\
                    & \mbox{or} & \nonumber        \\
W^*\rightarrow jj   &           & W\rightarrow l\nu. 
\end{eqnarray}
If the on-shell $W$ boson decays into leptons, then the $W$ mass constraint is used to fix the 
longitudinal component $p_L^\nu$ of the neutrino momentum. For this case, 
$W^*\rightarrow jj$ and all the jet-jet combinations are accepted. However, if 
$W^*\rightarrow l\nu$ instead, one can no longer use the $W$ mass constraint. In this case, 
$p_L^\nu$ is set to zero and the jet-jet combinations consistent with the $W$ mass are retained, 
i.e., $|m_W-m_{jj}| < 25$~GeV; in this mass window, the jet momenta are rescaled so that 
$m_{jj} = m_W$. Finally, we use the following $\chi^2$ criterion to select the best top quark, the neutral light Higgs and charged Higgs candidates:
\begin{eqnarray}
\label{eq:chi}
\chi^2 & = & (m_{Wb_i}-m_t)^2 + (m_{H^\pm b_l}-m_t)^2 \nonumber \\
       &   & + (m_{b_jb_k}-m_{h^0})^2.
\end{eqnarray}
Although the signal rate is initially two orders of magnitude smaller 
than the $t\bar{t}$ background rate as can be seen from Table~\ref{tab:who_1}, 
\begin{table*}
\begin{center}
\begin{minipage}{.6\linewidth} 
\caption{\label{tab:who_1} The expected rates ($\sigma\times BR$) for the signal 
$t\bar{t}\rightarrow bH^\pm Wb$ with \mbox{$H^\pm\rightarrow W^*h^0$} and for the $t\bar{t}$ 
backgrounds. It should be noted that due to the $\tan\beta$ dependence of the $t\rightarrow H^\pm 
b$ 
and of the $t\rightarrow Wb$ branching ratios, the $t\bar{t}$ background rates depend on 
$\tan\beta$.}
\end{minipage}
\begin{tabular}{lcccc} \hline
Process                           & $\tan\beta$ & $m_{h^0}$~(GeV) & $m_{H^\pm}$~(GeV) & 
$\sigma\,\times\,$~BR (pb) \\
\hline
$H^\pm\rightarrow W^*\,h^0$       &   2.0       &   83.5  & $152$      &    1.2                \\
                                  &   3.0       &   93.1  & $152$      &    0.2                \\
$t\bar{t}\rightarrow jjbl\nu b$   &   2.0       &         &            &    143                \\
                                  &   3.0       &         &            &    152                \\
 \hline 
\end{tabular}
\end{center}
\end{table*}
the proposed reconstruction procedure described 
in further detail in~\cite{ASSA3} permits the extraction of the signal with a significance exceeding $5\sigma$ in 
the low $\tan\beta\; (1.5-2.5)$ region as shown in Table~\ref{tab:who_3}. 
\begin{table*}
\begin{center}
\begin{minipage}{.7\linewidth} 
\caption{\label{tab:who_3} The expected signal-to-background ratios and significances for an 
integrated luminosity of 300~fb$^{-1}$. $\langle m_t\rangle$, $\langle m_{H^\pm}\rangle$ and 
$\langle m_{h^0}\rangle$ are the means of the Gaussian fits to the distributions of 
$m_{H^\pm b}$, 
$m_{W^*h^0}$ and $m_{b\bar{b}}$ respectively. The nominal values are shown in 
Table~\ref{tab:who_1} 
for $m_{H^\pm}$ and $m_{h^0}$. A central value of 175~GeV is taken for the top quark mass. 
$m_{H^\pm}$ and $m_t$ are not reconstructed at their nominal values (within the large statistical 
uncertainties, the numbers are consistent with the nominal values): this is due to the assumption 
that $p_L^\nu=0$ (made in selection cuts) and also to the fact that the 4-momentum of the $W^*$ 
in 
$H^\pm\rightarrow W^*h^0$ and $t\rightarrow W^*h^0b$ is not rescaled to the $W$ mass before 
reconstructing the charged Higgs and the top quark. The other 
top quark is reconstructed to the nominal value (see~\cite{ASSA3} for details) since here, 
$t\rightarrow Wb$; this $W$ boson is on-shell: 
in the leptonic channel the $W$ mass constraint guarantees that $m_{l\nu}=m_W$, and in the 
hadronic channel, the jet momenta are rescaled within a mass window so that $m_{jj}=m_W$.  
The 
significances and the signal-to-background ratios are calculated within $\pm 2\sigma_{H^\pm}$ of 
$\langle m_{H^\pm}\rangle$.}
\end{minipage}
\begin{tabular}{lcc} \hline
                                 & $\tan\beta=2$  & $\tan\beta=3$ \\
\hline
$\langle m_t\rangle$~(GeV)        & $188\pm 20$    & $190\pm 29$   \\
$\sigma_t$~(GeV)                 & $18\pm 11$     & $20\pm 10$    \\
$\langle m_{H^\pm}\rangle$~(GeV)  & $157\pm 7$     & $160\pm 10$   \\
$\sigma_{H^\pm}$~(GeV)           & $19\pm 8$      & $21\pm 10$    \\
$\langle m_{h^0}\rangle$~(GeV)    & $83\pm 1$      & $92\pm 4$      \\
$\sigma_{h^0}$~(GeV)             & $12\pm 1$      & $13\pm 3$      \\ 
Signal events                    & 136            & 25               \\
Background events                & 40             & 43              \\
$S/B$                            & 3.4            & 0.6            \\
$S/\sqrt{B}$                     & 21.5           & 3.8              \\
 \hline
\end{tabular}
\end{center}
\end{table*}
At high $\tan\beta$, though the reconstruction efficiency 
remains comparable, the signal rate decreases so significantly that the discovery potential 
vanishes in this region.

\subsection{$\bm{H^\pm\rightarrow W^\pm h^0}$, $\bm{m_{H^\pm} > m_t}$}

Above the top quark mass, the charged Higgs is produced in association with a top quark according 
to:
\begin{eqnarray}
gb & \rightarrow & tH^\pm \nonumber \\
H^\pm & \rightarrow & Wh^0 \nonumber \\
t & \rightarrow & Wb.
\end{eqnarray}
The final state for the signal contains two $W$ bosons, one of which is required to decay into 
leptons (electron or muon to trigger the experiment), the other into jets, and three $b$-jets due 
to the subsequent decay $h^0\rightarrow b\bar{b}$. The background in this case comes from 
$t\bar{t}b$ and $t\bar{t}q$ events with both top quarks decaying into $W$'s. 
We search for one isolated lepton, three $b$-tagged jets and at least two non $b$-jets. In this case, both
$W$-bosons are on the mass shell. The $W$ mass constraint is used to find a longitudinal component 
of the neutrino momentum, $W^\pm\to l\nu$, and the jet-jet combinations satisfying the 
requirement 
$|m_{jj}-m_W| < 25$~GeV are retained as done in the previous analysis. The associated top 
quark from $gb \to tH^\pm$, $t \to W_ib_k$ and the neutral light Higgs, $h^0 \to b_lb_m$ are 
reconstructed by minimising $\chi^2 = (m_{W_ib_k}-m_t)^2 + (m_{b_lb_m}-m_{h^0})^2$. Initially, 
the total background is at least three orders of magnitude higher than the signal in the most 
favourable case ($\tan\beta=3$), as shown in Table~\ref{tab:who_4}. 
\begin{table*}
\begin{center}
\begin{minipage}{.6\linewidth} 
\caption{\label{tab:who_4} The rates for the signal $bg\rightarrow H^\pm t\rightarrow Wh^0\,Wb$ 
and the $t\bar{t}$ background as a function of $\tan\beta$.}
\end{minipage}
\begin{tabular}{lcccc} \hline
Process                           & $\tan\beta$ & $m_{h^0}$~(GeV) & $m_{H^\pm}$~(GeV) & 
$\sigma\,\times\,$~BR (pb) \\
\hline
$H^\pm\rightarrow W^\pm\,h^0$         &   1.5       &   78.0  & $250$      & 0.023                   \\
                                  &   3.0       &   99.1  & $200$      & 0.134                   \\
                                  &   5.0       &   104.9 & $200$      & 0.031                   \\
$t\bar{t}\rightarrow jjbl\nu b$   &             &         &            & 228                     \\
 \hline 
\end{tabular}
\end{center}
\end{table*}
However, with the reconstruction technique described in detail  
in~\cite{ASSA3}, the signal-to-background ratios could be improved by two orders of magnitude. 
This improvement is still insufficient to observe the signal over the background; for example, 
for $\tan\beta=3$ and $m_{H^\pm}=200$~GeV, a significance of only 3.3 can be expected after 
3 years at high luminosity. A parton level study of this channel was carried out in~\cite{Wh0} 
using $gg \to tbH^\pm$ and a viable signal survives above the main irreducible $tbW^\pm h^0$ 
continuum for $\tan\beta \sim 2-3$ and $m_{H^\pm} \sim 200$~GeV, in fair agreement with the current 
analysis.

The main objective of this study is to  demonstrate a good signal reconstruction and a high 
background suppression with the ATLAS detector. Indeed, although the
signal is marginally  viable in MSSM, the results can be normalised to other
models, for instance NMSSM, where LEP  constraints no longer apply and the
discovery potential in this channel would extend to a  significant area of the
parameter space as explained below.  

\subsection{\bm{$H^\pm\rightarrow W^\pm h^0$} in NMSSM}

In the MSSM, the relation~(\ref{eq:mA_mHp}) and the direct lower limit on the CP-odd Higgs boson 
mass from LEP (see the introduction section) translate into an indirect lower bound on the 
charged Higgs, $m_{H^\pm} \OOrd 120$~GeV; in fact for $\tan\beta\sim 3$, $m_{H^\pm} \OOrd 
250$~GeV~\cite{MOR-WH03,LEP-SM}. As a result, the channel $H^\pm\rightarrow W^\pm h^0$ has a high 
threshold in the LEP allowed region where the branching ration is very small (see 
Fig.~\ref{fig:HDECAYS}). Indeed,  beyond $\tan\beta=3$, as demonstrated by the study shown here, 
this channel presents no discovery potential due to the very low signal rate. It has been argued 
that in the singlet extension to MSSM, i.e., NMSSM, this channel is immune to the LEP constraints 
and there may be a significant discovery potential above and below the top quark 
mass~\cite{MOR-WH03}. In fact, NMSSM extends the Higgs sector of the MSSM by 
adding a complex singlet scalar field and seven physical Higgs bosons are predicted in this model, 
three neutral CP-even $h^0$, $H^0_1$, and $H^0_2$, two CP-odd $A^0_1$ and $A^0_2$, and a charged 
pair $H^\pm$~\cite{2HDM}. The parameter space is 
therefore less constrained than the one of the MSSM and the indirect lower
limits on the  Higgs masses from LEP are no longer valid. In addition, the
mixing between the  singlet and the doublet states would dilute the direct mass
limits on the latter from LEP.  Consequently, the channel $H^\pm\rightarrow
W^\pm\, (h^0,\,A^0)$ can be the dominant decay mode  for low $\tan\beta$ and
$m_{H^\pm} \sim 160$~GeV. Therefore, The $H^\pm\rightarrow W^\pm h^0$  channel
which is marginally viable in MSSM would yield a significant signal in 
NMSSM~\cite{MOR-WH03}.

\subsection{\bm{$H^\pm \to W^\pm\,H^0(A^0)$}}

It is shown in~\cite{Akero1} that large mass splittings ($m_{H^\pm} > m_{H^0} \OOrd 100$~GeV) between 
the charged Higgs and the heavier CP-even Higgs bosons are possible in MSSM if the $\mu$ parameter 
is considerably larger than the common SUSY scale $M_{SUSY}$. The decay $H^\pm \to W^\pm H^0$ would then 
have a sizeable branching ratio, thus offering detection prospects in the MSSM parameter space where 
$\mu$ is large.

In the general 2HDM, the decays $H^\pm \to W^\pm\,H^0(A^0)$ can be dominant in a wide range of the 
parameter space accessible at present and future colliders~\cite{Akero2}.

\section{$\bm{m_{H^\pm}}$ and $\bm{\tan\beta}$ Determination}
In this section, we discuss the expected precisions on the charged
Higgs mass and $\tan\beta$ measurements with the ATLAS detector in the
$H^\pm\rightarrow tb$ and $H^\pm\rightarrow\tau^\pm\nu_\tau$ channels.

\subsection{$\bm{H^\pm}$ mass determination in $\bm{H^\pm\rightarrow\tau^\pm\nu_\tau}$}

As discussed in section~\ref{sec:taunu}, this channel does not offer the possibility for the 
observation of a resonance peak above the background, only the transverse Higgs mass can be 
reconstructed because of the presence of the neutrino in the final state. The background comes 
from single top ($Wt$) and $t\bar{t}$ productions with one $W^\pm\rightarrow\tau^\pm\nu_\tau$. 
Thus, the transverse mass is kinematically constrained to be less than the $W$-boson mass while 
in the signal, the upper bound is the charged Higgs mass. 
\begin{table*}  
\begin{center}
\begin{minipage}{.65\linewidth}
\caption{\label{tab:mass_2} The systematic effects on the mass determination in the 
$H^\pm\rightarrow\tau^\pm\nu_\tau$ channel are small. Columns 2 and 3 show the statistical 
uncertainties for an integrated luminosity of 300~fb$^{-1}$. Columns 4 and 5 include the 
systematic uncertainties. The total uncertainties are dominated by the statistical errors.}
\end{minipage}
\begin{tabular}{ccccc}\hline
$m_{H^\pm}$~(GeV) & \multicolumn{2}{c}{Statistics only} & \multicolumn{2}{c}{With systematics} \\ 
\hline
       & $\langle m\rangle$   & $\delta m$ & $\langle m\rangle$  & $\delta m$ \\
225.9  & 226.4  & 1.7 & 225.9  & 1.7 \\
271.1  & 271.1  & 2.0 & 270.9  & 2.3 \\
317.8  & 318.3  & 3.0 & 319.9  & 3.5 \\
365.4  & 365.7  & 4.6 & 365.2  & 4.7 \\
413.5  & 413.8  & 4.5 & 414.9  & 4.7 \\
462.1  & 462.6  & 6.0 & 460.8  & 6.3 \\
510.9  & 511.9  & 7.4 & 511.7  & 9.2 \\ 
\hline
\end{tabular}
\end{center}
\end{table*}

The differences in the event topology and in the $\tau$ polarisation have been used to suppress 
the backgrounds as discussed in section~\ref{sec:taunu}~\cite{ASSA2,RITVA}, so that above the 
$W$ mass threshold, the background in this channel is relatively small as shown in 
Fig.~\ref{fig:taunu5}. As a result, although there is no reconstruction of the resonance peak in 
this channel, the Higgs mass can be extracted from the transverse mass distribution with a 
relatively good precision. For the mass determination in this channel, we use the likelihood method presented in~\cite{HOHL,ASSA4}.

Three main sources of systematic uncertainties are included in the mass determination: the shape 
of the background, the background rate and the energy scale. The background shape becomes more 
significant at lower Higgs masses where there is more overlap between signal and background. To 
include this effect, we assumed a linear variation of the background shape, from $-10$\% to 
$+10$\% between the minimum and the maximum of the transverse mass distribution. Another source 
of systematic uncertainty is the rate of the backgrounds. It is expected that the background 
rate ($Wt$ and $t\bar{t}$) could be known to 5\%~\cite{HOHL}. Therefore, to take this 
effect 
into account, we increase the background rate by 5\% while at the same time we decrease the 
signal by 5\%. Finally, we also include the scale uncertainty: 1\% for jets and 0.1\% for 
photons, electrons and muons. In Table~\ref{tab:mass_2}, we show the effects of the systematic
uncertainties: the overall uncertainty in the mass determination is dominated by statistics.

The overall relative precision in this channel ranges from 1.3\% at $m_{H^\pm}=226$~GeV to 3.1\% 
at $m_{H^\pm}=511$~GeV for an integrated luminosity of 100 fb$^{-1}$. At 300~fb$^{-1}$, the 
precision improves to 0.8\% at $m_{H^\pm}=226$~GeV and 1.8\% at $m_{H^\pm}=511$~GeV~\cite{ASSA4}.
 
\subsection{$\bm{H^\pm}$ Mass Determination in $\bm{H^\pm\rightarrow tb}$}

In the $tb$ channel, the full invariant mass can be reconstructed as shown in 
Fig.~\ref{fig:tb_sig6} although this channel suffers from the large irreducible $t\bar{t}b$ 
background and also from the signal combinatorial background. The determination of the mass can 
be done using the likelihood method described in~\cite{ASSA4,HOHL} or by fitting the signal and the background. 
In the latter case, one assumes that the background shape and normalisation can be determined by 
fitting outside the signal region, thus, the systematic uncertainties include only the scale 
uncertainty. We assume a Gaussian shape for the signal and an exponential for the background and 
fit signal$+$background including the statistical fluctuations and the scale uncertainty. The 
precisions on the mass determination from the likelihood and fitting methods are comparable.

The relative precision in this channel ranges from 0.8\% at
$m_{H^\pm}=226$~GeV to 5.2\% at $m_{H^\pm}=462$~GeV for 100~fb$^{-1}$. For
300~fb$^{-1}$, the precision improves to 0.5\% at 226~GeV and 3.5\% at
462~GeV~\cite{ASSA4}.

\subsection{Determination of $\bm{\tan\beta}$}
\label{detbeta}

It is possible to determined $\tan\beta$ by measuring the signal rate in the $\tau\nu$ channel 
where the backgrounds are relatively low. The main systematic error would come from the knowledge 
of the luminosity. The uncertainty in the rate measurement can be estimated as~\cite{GUNI}:
\begin{equation}
\label{eq:rate}
\frac{\Delta (\sigma\times BR)}{\sigma\times BR} = \sqrt{\frac{S+B}{S^2} + \left(\frac{\Delta 
\mathcal{L}}{\mathcal{L}}\right)^2},
\end{equation}
where the relative uncertainty on the luminosity measurement is expected to be $\sim$5\%. 
The uncertainty on $\tan\beta$ is computed as:
\begin{equation}
\label{eq:dtan}
\Delta\tan\beta \simeq \Delta(\sigma\times BR)\left[\frac{d(\sigma\times BR)}{d\tan\beta}\right]^{-1}.
\end{equation}
At large $\tan\beta$,  from Equations~(\ref{gbcross}) and~(\ref{eq:BR}), the rate in the 
$H^\pm\rightarrow\tau^\pm\nu_\tau$ channel is obtained as:
\begin{equation}
\label{eq:sigbr}
\sigma\times BR \propto \tan^2\beta.
\end{equation}
From the relations~(\ref{eq:dtan}) and~(\ref{eq:sigbr}), we get:
\begin{equation}
\label{eq:errtan}
\frac{\Delta\tan\beta}{\tan\beta} = \frac{1}{2}\frac{\Delta(\sigma\times BR)}{\sigma\times BR}.
\end{equation}
The relative precision on $\tan\beta$ ranges from 15.4\% to 7.3\% for $\tan\beta=20$ to 50, at 
low luminosity. For an integrated luminosity of 300~fb$^{-1}$, the precision improves to 7.4\% 
at $\tan\beta=20$ and to 5.4\% at $\tan\beta=50$~\cite{ASSA4}. Theoretical uncertainties of 8\% 
are also included in the estimates of the expected relative precision on $\tan\beta$.
\begin{figure}
\epsfxsize=8truecm
\begin{center}
  \epsffile{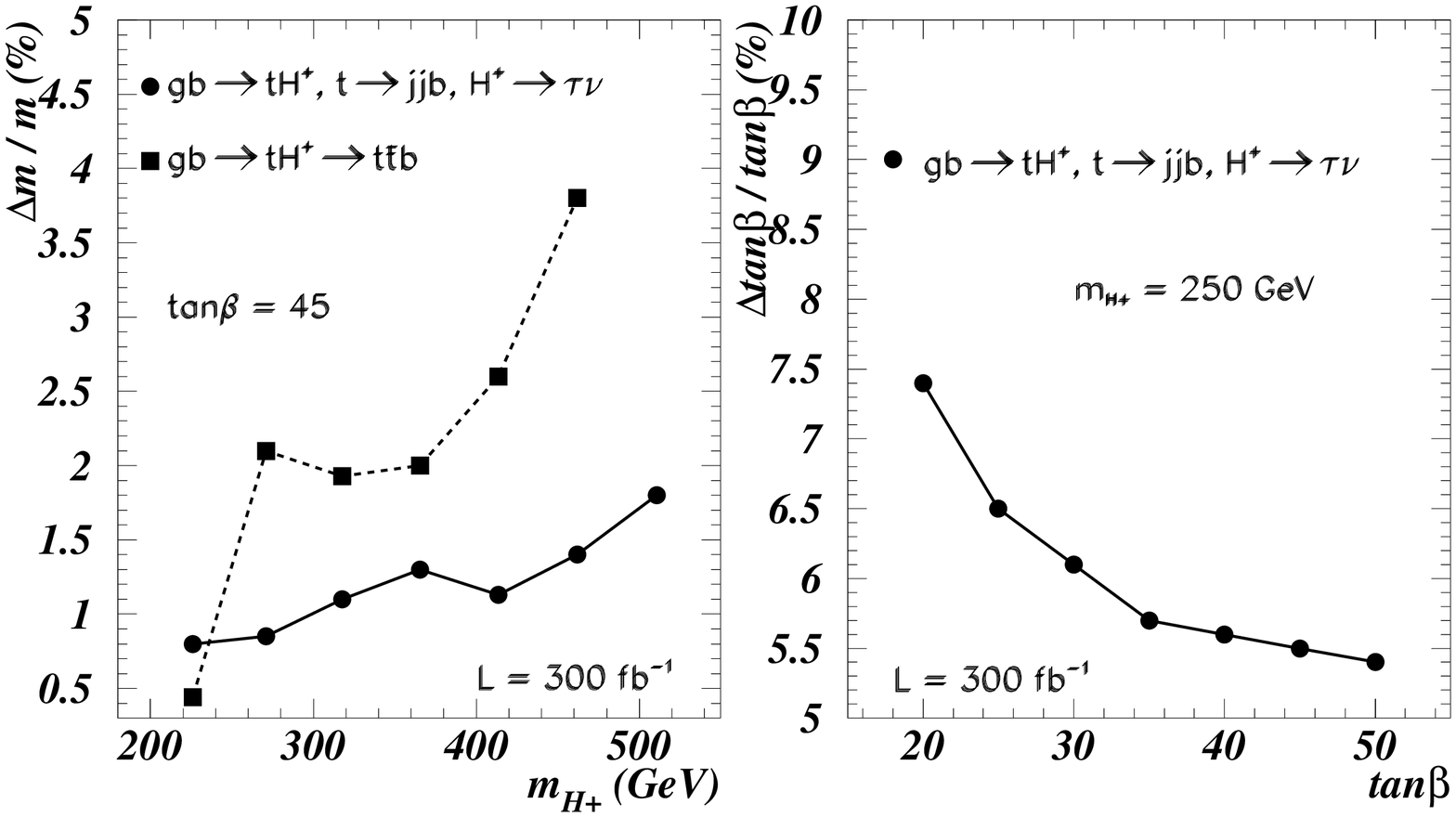}
\caption{The expected overall precision on the charged Higgs mass and on $\tan\beta$ 
measurements, as a function of the charged Higgs mass (left plot) and $\tan\beta$ (right plot) 
respectively. For the mass determination, the $H^\pm\rightarrow\tau^\pm\nu_\tau$ channel gives 
better precisions than $H^\pm\rightarrow tb$ except at low Higgs masses. In addition, 
$H^\pm\rightarrow\tau^\pm\nu_\tau$ allows for the determination of $\tan\beta$ by measuring the 
rate in this channel.}
\label{fig:mass5}
\end{center}
\end{figure}

\begin{table*}
 \begin{center}
\begin{minipage}{.72\linewidth}
\caption{\label{tab:mass_6} The overall precisions on the mass determination are better in the 
$\tau\nu$ channel than in the $tb$ channel. This is due to the fact that the latter suffers from 
large $t\bar{t}b$ and signal combinatorial backgrounds ($\mathcal{L}=100$~fb$^{-1}$).}
\end{minipage}
\begin{tabular}{ccccc}\hline
$m_{H^\pm}$~(GeV) & \multicolumn{2}{c}{$H^\pm\rightarrow\tau^\pm\nu_\tau$} & 
\multicolumn{2}{c}
{$H^\pm\rightarrow tb$} \\ \hline
      & $\langle m\rangle$   & $\delta m$ & $\langle m\rangle$ & $\delta m$ \\
225.9 & 225.9 & 2.9 & 226.9 & 1.8 \\
271.1 & 271.0 & 3.9 & 270.1 & 10.1 \\
317.8 & 319.7 & 5.9 & 320.2 & 11.3 \\
365.4 & 364.9 & 8.1 & 365.4 & 12.1 \\
413.5 & 414.8 & 8.0 & 417.4 & 17.6 \\
462.1 & 460.7 & 10.6 & 465.9 & 24.1 \\
510.9 & 511.4 & 15.7 &             &           \\ 
\hline
\end{tabular}
\end{center}
\end{table*}

Fig.~\ref{fig:mass5} illustrates the expected overall precision on the charged Higgs mass and 
$\tan\beta$ determination for an integrated luminosity of 300~fb$^{-1}$. In either channel, the 
overall uncertainties are dominated by the statistical errors. The $\tau\nu$ channel offers 
better precisions on the Higgs mass determination than the $tb$ channel, except at low
Higgs masses where the $\tau\nu$ channel suffers from a much reduced selection
efficiency or a much higher background level as shown in Table~\ref{tab:mass_6} and 
Fig.~\ref{fig:mass5}~\cite{ASSA4}. 

\section{$\bm{H^\pm\rightarrow\tau^\pm\nu_\tau}$ in Large Extra Dimensions}  
\label{sec:extra}

In models where extra dimensions open up at the TeV scale, small neutrino masses can be generated 
without implementing the seesaw mechanism \cite{KK,1,2}. These models postulate the existence of 
$\delta$ additional spatial dimensions of size $R$ where gravity and perhaps 
other fields freely propagate while the SM degrees of freedom are confined to 
(3+1)--dimensional wall (4D) of the higher dimensional space. The idea that our
world could be a topological defect of a higher--dimensional theory~\cite{rub}
finds a natural environment in string theory~\cite{pol}.
 
The right--handed neutrino can be interpreted as a singlet with no quantum 
numbers to constrain it to the SM brane and thus, it can propagate into the 
extra dimensions just like gravity~\cite{3}. Such singlet states in the bulk 
couple to the SM states on the brane as right--handed neutrinos with small 
couplings --- the Yukawa couplings of the bulk fields are suppressed by the 
volume of the extra dimensions. The interactions between the bulk neutrino  
and the wall fields generate Dirac mass terms between the wall fields and all 
the Kaluza-Klein modes of the bulk neutrino. As long as this mass is less than 
$1/R$, the Kaluza-Klein modes are unaffected while for the zero mode, the 
interaction generates a Dirac neutrino mass suppressed by the size of the 
extra dimensions:  
\begin{equation}  
\label{eq:dm}  
m_D =\frac{\lambda}{\sqrt{2}}\frac{M_*}{M_{Pl}}v  
\end{equation}  
where $\lambda$ is 
a dimensionless constant, $v$ the Higgs vacuum expectation value (see the introduction 
section), $M_{Pl}=2.4 \times 10^{18}$ GeV is the reduced Planck scale related to the 
usual Planck mass $1.2 \times 10^{19}$ GeV  $=\sqrt{8\pi} M_{Pl}$, and $M_*$ is the true 
scale of gravity, or the fundamental Planck scale of the ($4+\delta$)D space time:   
\begin{equation} 
\label{eq:scale}  
M_{Pl}^2 = R^\delta M_*^{\delta+2}.
\end{equation}  
The mixing between the lightest neutrino with mass $m_D$
and the  heavier neutrinos introduces a correction $N$ to the Dirac mass such
that the  physical neutrino mass $m_\nu$ is~\cite{2}:  
\begin{equation}   
\label{eq:nu}  
m_\nu =\frac{m_D}{N},   
\end{equation}   
where  
\begin{equation}  
\label{eq:N_prime} N  
\simeq 1 + \left(\frac{m_D}{M_*}\right)^2\left(\frac{M_{Pl}}{M_*}\right)^2 
\frac{2\pi^{\delta/2}}{\Gamma(\delta/2)}\frac{1}{\delta-2} \; .
\end{equation} 
As shown in Table~\ref{tab:table1}, small neutrino masses, $m_\nu$, can be obtained 
consistent with atmospheric neutrino oscillations~\cite{4}.  
\begin{table*} 
\begin{center}  
\begin{minipage}{.92\linewidth} 
\caption{\label{tab:table1}The parameters used in the current analysis of the 
signal with the corresponding polarisation asymmetry. In general, $H^-$ would 
decay to $\tau^-_L$ and $\tau^-_R$, $H^-\rightarrow\tau_R^-\bar{\nu} + 
\tau_L^-\psi$, depending on the asymmetry. For the decay 
$H^-\rightarrow\tau^-_R\bar{\nu}$ (as in MSSM), the asymmetry is $-1$ and this case is 
already studied for the LHC~\cite{ASSA2,RITVA} and discussed in section~\ref{sec:taunu}. The signal 
to be studied is $H^-\rightarrow\tau^-_L\psi$.}  
\end{minipage}
\vbox{\offinterlineskip 
\halign{&#& \strut\quad#\hfil\quad\cr    
\colrule 
& &&$M_*$ (TeV) && $\delta_\nu$ 
&& $\delta$ && $m_D$ (eV) &&  $m_{H^\pm}$ (GeV) && $\tan\beta$ && Asymmetry  
&& $m_\nu$ (eV) &\cr 
\colrule  
&Signal-1 && 2 && 4 && 4 && 3.0  &&  219.9 &&  30  && $\sim 1$ && 0.5 $10^{-3}$ 
\cr   
&Signal-2 && 20 && 3 && 3 && 145.0 && 365.4 && 45 && $\sim 1$ && 0.05 &\cr 
&Signal-3 && 1 && 5 && 6 && 5.0 && 506.2 && 4 && $\sim 1$ && 0.05 &\cr   
&Signal-4 && 100 && 6 && 6 && 0.005 && 250.2 && 35&& $\sim -1$ && 0.005 &\cr 
&Signal-5 && 10 && 4 && 5 && 0.1 && 350.0 && 20 && $\sim -1$ && 0.04 &\cr  
&Signal-6 && 50 && 5 && 5 && 0.04 && 450.0 && 25 && $\sim -1$ && 0.04 &\cr  
\colrule
}}   
\end{center}  
\end{table*} 

$H^-$ decays to the right--handed 
$\tau^-$ through the $\tau$ Yukawa coupling:     
\begin{equation}  
\label{eq:mssm}   
H^-\rightarrow \tau_R^-\bar{\nu}.   
\end{equation}  
The $H^-$ decay to left--handed $\tau^-$ is completely suppressed in MSSM. 
However, in the scenario of singlet neutrino in large extra dimensions, $H^-$ 
can decay to both right--handed and left--handed $\tau^-$ depending on the 
parameters $M_*$, $m_D$, $\delta$, $m_{H^\pm}$ and $\tan\beta$ (see
\cite{ASSA5} for detailed formulas):  
\begin{equation}  
\label{eq:extraD}  
H^- \rightarrow \tau_R^-\bar{\nu} +\tau_L^-\psi,  
\end{equation}  
where $\psi$ is a bulk neutrino and $\nu$ is 
dominantly a light neutrino with a small admixture of the Kaluza-Klein modes.
The measurement of the polarisation asymmetry,   
\begin{equation}  
\label{eq:asym}  
A =\frac{\Gamma(H^-\rightarrow\tau_L^-\psi)- 
\Gamma(H^-\rightarrow\tau_R^-\bar{\nu})} 
{\Gamma(H^-\rightarrow\tau_L^-\psi)+ 
\Gamma(H^-\rightarrow\tau_R^-\bar{\nu})}, 
\end{equation}  
can be used to distinguish between the ordinary 2HDM-II and the 
scenario of singlet neutrino in large extra dimensions. In the 2HDM-II, the
polarisation asymmetry would be  $-1.0$. In the framework of large extra
dimensions, the polarisation  asymmetry can vary from $+1$ to $-1$. In the
latter case, the decay  of $H^-$ is similar to the 2HDM-II but
possibly with a different  phase space since the neutrino contains some
admixture of the Kaluza-Klein  modes.    
The singlet neutrino may not necessarily propagate into the 
$\delta$-extra dimensional space. It is possible to postulate that the singlet 
neutrino propagate into a subset $\delta_\nu$ ($\delta_\nu \leq \delta$) of 
the $\delta$ additional spatial dimensions, in which case the formalism for 
the generation of small Dirac neutrino masses is merely a generalisation of 
the case $\delta_\nu=\delta$ discussed above~\cite{1}. 
 
The charged Higgs decay to right--handed $\tau$, $H^-\rightarrow\tau_R^-\bar{\nu}$ have 
been extensively studied for the LHC as discussed above~\cite{ASSA2,RITVA}. In this section, we 
discuss the possibility to observe $H^-\rightarrow\tau_L^-\psi$ at the LHC above the top quark 
mass \cite{agashe,ASSA5}. Table~\ref{tab:table1} shows the parameters selected
for the  current analysis. The cases  where the asymmetry is $+1$ are
discussed. No additional Higgs bosons are needed. As a result, the
charged Higgs production mechanisms  are the same as in the 2HDM-II as shown
in Fig.~\ref{fig:prod_graphs}.   We consider the $2\rightarrow 2$ production 
process where the charged Higgs is produced with a top quark, $gb\rightarrow 
tH^\pm$. Further, we require the hadronic decay of the top quark, 
$t\rightarrow Wb\rightarrow jjb$ and the charged Higgs decay to 
$\tau$ leptons. 
 
The major backgrounds are the single top production 
$gb\rightarrow Wt$, and $t\bar{t}$ production with one $W^+\rightarrow jj$ and 
the other $W^-\rightarrow\tau_L^-\bar{\nu}$ (there is no enhancement in the background rate 
from the contribution $W^- \to \tau^-_L\psi$). Depending on the polarisation 
asymmetry --- Equation~(\ref{eq:asym}) --- $H^-\rightarrow\tau_R^-\bar{\nu}$ will 
contribute as an additional background. In Table~\ref{tab:table2}, we list the 
rates for the signal and for the backgrounds.  
\begin{table*} 
\begin{center}  
\begin{minipage}{.78\linewidth} 
\caption{\label{tab:table2}The expected rates ($\sigma\times$ BR), for the signal 
$gb\rightarrow t H^\pm$  with $H^-\rightarrow\tau_R^-\bar{\nu}+\tau_L^-\psi$ 
and $t\rightarrow jjb$, and for the backgrounds:  $W t$ and $t\bar{t}$ 
with $W^-\rightarrow\tau_L^-\bar{\nu}$ and $W^+\rightarrow jj$. We assume an 
inclusive $t\bar{t}$ production cross section of 590~pb. Other cross  sections 
are taken from PYTHIA~6.1. See  Table~\protect{\ref{tab:table1}} for the 
parameters used for Signal-1, Signal-2 and Signal-3. In the  last columns, we 
compare the $H^\pm\rightarrow\tau^\pm\nu_\tau$ branching ratios in this model 
to the corresponding MSSM branching ratios.} 
\end{minipage} 
\vbox{\offinterlineskip 
\halign{&#& \strut\quad#\hfil\quad\cr 
\colrule 
&Process && $\sigma\,\times\,$~BR (pb) && BR($H^\pm\rightarrow\tau\nu+\tau\psi$) &&  
MSSM: BR($H^\pm\rightarrow\tau^\pm\nu_\tau$)&\cr  
\colrule  
&Signal-1 && 1.56 && 0.73 &&  0.37  &\cr 
&Signal-2 && 0.15 && 1.0 && 0.15 & \cr  
&Signal-3 && 0.04 && 1.0  &&   0.01 & \cr  
\colrule 
&$t\bar{t}$ && 84.11 &&  &&   &  \cr  
&$gb\rightarrow Wt$ ($p_T>30$~GeV) && 47.56 &&  &&   &  \cr 
\colrule}} 
\end{center}  
\end{table*} 
 
In general, $H^-\rightarrow\tau_L^-\psi+\tau_R^-\bar{\nu}$ 
with the asymmetry between -1 and 1~\cite{agashe}. However, the study of 
$H^-\rightarrow\tau_R^-\bar{\nu}$ has been carried out in detail and discussed in 
section~\ref{sec:taunu}. Therefore, in the current study, we consider the 
parameters shown in Table~\ref{tab:table1} and Table~\ref{tab:table2} for 
which the asymmetry is one, i.e., $H^-\rightarrow\tau_L^-\psi$.  
  
The polarisation of the $\tau$ lepton is included in this analysis through  
TAUOLA~\cite{TAUOLA}. We consider the hadronic one-prong decays of the 
$\tau$ lepton  --- see the relations~(\ref{eq:pinu}). For the signal in the MSSM, 
right--handed $\tau_R^-$'s come from the charged Higgs decay, $H^-\rightarrow\tau_R^-\bar{\nu}$, 
while in the backgrounds, left--handed $\tau_L^-$'s come from the decay of the 
$W^-(\rightarrow\tau_L^-\bar{\nu})$. 
\begin{figure} 
\epsfysize=8truecm 
\begin{center} 
\epsffile{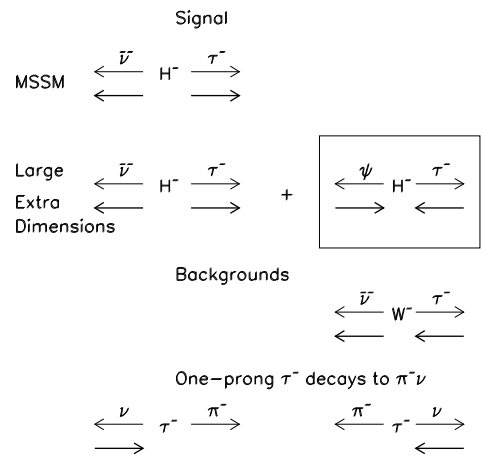} 
\caption{Polarisation of the decay $\tau$ from $H^\pm$ in the MSSM and in models with  
a singlet neutrino in large extra dimensions. In the latter case, both left 
and right--handed $\tau$'s can be produced with some polarisation asymmetry. In 
the backgrounds, the $\tau$ comes from the decay of the $W^\pm$. The signal to 
be studied is in the box --- the polarisation of the decay $\tau$ in this 
signal is the same as in the background. Thus, $\tau$ polarisation effects 
would not help in suppressing the backgrounds but they may help distinguish 
between the 2HDM and other models.}  
\label{fig:tau_pol_led}  
\end{center}
\end{figure}
In the MSSM, the requirement~(\ref{eq:pfrac}) would retain 
only the $\pi$ and half of the longitudinal $\rho$ and $a_1$ contributions while eliminating 
the transverse components along with the other half of the longitudinal contributions. In 
addition, this requirement would suppress much of the backgrounds. In the framework of 
large extra dimensions, we are interested in $H^-\rightarrow\tau_L^-\psi$ 
where, as shown in Fig.~\ref{fig:tau_pol_led}, the polarisation of the 
$\tau$ lepton would be identical to the background case but opposite to the 
MSSM case. Therefore, the requirement~(\ref{eq:pfrac}) would not help in 
suppressing the backgrounds. Nevertheless, there are still some 
differences in the kinematics which can help reduce the background level, and 
the selection criteria are similar to the case presented in section~\ref{sec:taunu}, except 
here, we search for one-prong hadronic $\tau$ decays (see~\cite{ASSA5} for 
further details):  
\begin{description} 
\item[1.] The missing transverse momentum and the momentum of the 
$\tau$ jet are increasingly harder as the charged Higgs mass increases. 
 
\item[2.] The difference in the azimuthal opening angle between the 
$\tau$ jet and the missing transverse momentum explained in section~\ref{sec:taunu}.  
 
\item[3.] The difference in the transverse mass --- Equation(~\ref{eq:trans}) --- 
between the signal $H^-\rightarrow\tau^-_L\psi$ and the background 
$W^-\rightarrow\tau^-_L\bar{\nu}$.
\end{description}
 
\begin{figure} 
\epsfysize=8truecm 
\begin{center} 
\epsffile{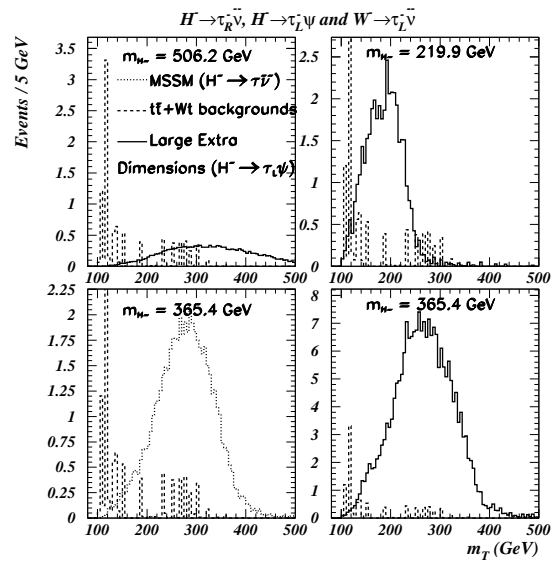} 
\caption{The reconstructions of the transverse mass for the signal in the MSSM, the signal
in models with a singlet neutrino in large extra dimensions and for the 
backgrounds, for an integrated luminosity of 100~fb$^{-1}$. In general, an MSSM charged 
Higgs can be discovered at the LHC  depending on $m_A$ and $\tan\beta$. In the 
models with a singlet neutrino in large extra dimensions, the signal can also 
be discovered at the LHC depending on the parameters $M_*$, $\delta$, $m_D$, 
$m_A$ and $\tan\beta$. The observation of the signal in the transverse mass 
distribution would not be sufficient to identify the model: the $\tau$ 
polarisation effects must be explored further.}  
\label{fig:led_tnu_bgd} 
\end{center}  
\end{figure}

\begin{figure} 
\epsfysize=8truecm 
\begin{center} 
\epsffile{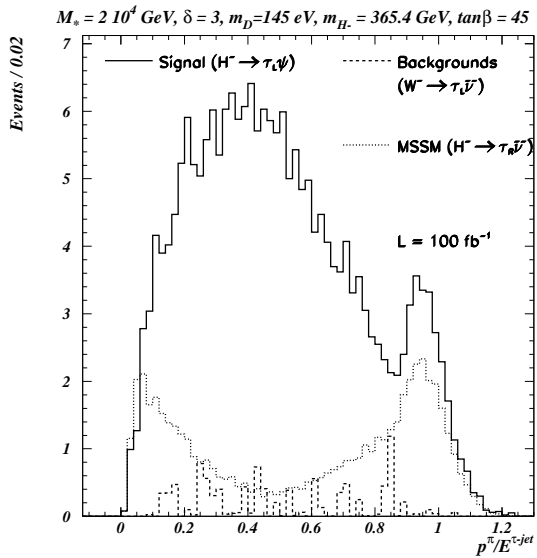} 
\caption{The distribution of the ratio of the charged pion track momentum in one  
prong $\tau$ decay to the $\tau$-jet energy for $m_A=350$~GeV, 
$\tan\beta=45$, $M_*= 20$~TeV, $\delta=3$ and $m_\nu=0.05$~eV. In the 2HDM-II, 
this ratio would peak near 0 and 1 as shown while in other models, the actual 
distribution of this ratio would depend on the polarisation asymmetry since 
both left and right--handed $\tau$'s would contribute. In the case shown, the 
asymmetry is $\sim 1$ and the ratio peaks near the centre of the 
distribution.} \label{fig:350_45_145_1}  
\end{center}  
\end{figure}

The reconstruction of the transverse mass (see 
Fig.~\ref{fig:led_tnu_bgd}) is not enough to distinguish between the MSSM and 
the singlet neutrinos in large extra dimensions. 
The differences in these two scenarios are best seen in the distribution of 
$p^\pi/E^{\tau-jet}$, the fraction of the energy carried by the charged track which is 
shown in Fig.~\ref{fig:350_45_145_1}.  

The mass of the neutrino $\psi$ would be different on an event by event basis. Consequently, 
the efficiencies of the kinematic cuts would be somewhat different. However, the main results of 
the current analysis derive from the differences in the polarisations of the $\tau$ lepton and 
in the transverse mass bounds, and would not be significantly affected by the neutrino mass effect.

Although the observation of a signal in the transverse mass 
distribution can be used to claim discovery of the charged Higgs, it is 
insufficient to pin down the scenario that is realized. Additionally, by 
reconstructing the fraction of the energy carried  by the charged track in the 
one-prong $\tau$ decay, it is possible to claim whether the scenario is the 
ordinary 2HDM or not. The further measurement of the polarisation asymmetry 
might provide a distinctive evidence for models with singlet neutrinos in large 
extra dimensions. 

\section{$\bm{H^\pm}$ Discovery Potential}

\begin{figure}
\epsfxsize=8truecm
\begin{center}
\epsffile{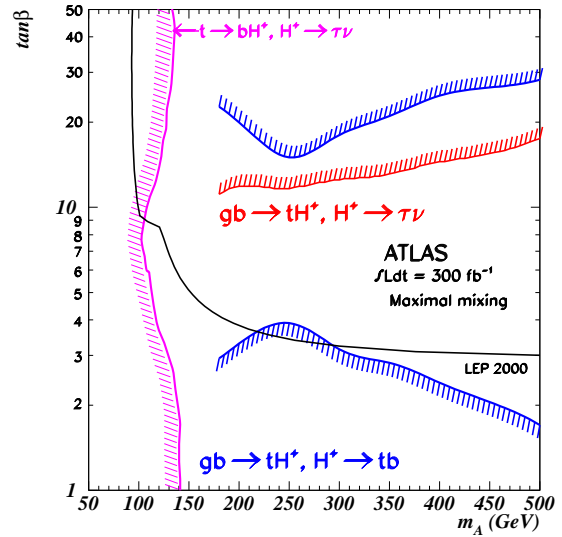}
\caption{\small The ATLAS 5-$\sigma$ discovery contour of the charged Higgs. Below the top quark 
mass, the charged Higgs is produced from top decay and the $\tau^\pm\nu_\tau$ channel provides 
coverage for most $\tan\beta$ below $\sim$160~GeV. Above the top quark mass, the $tb$ channel 
covers the low and the high $\tan\beta$ regions while the $\tau^\pm\nu_\tau$ channel extends the discovery 
reach to high Higgs mass and to lower $\tan\beta$ in the high $\tan\beta$ region.}
\label{fig:contour}
\end{center}
\end{figure}

In the $H^\pm\rightarrow tb$ channel, upwards of 5-$\sigma$ discovery can be achieved above the 
top quark mass in the low and high $\tan\beta$ regions up to $\sim$400~GeV as discussed 
above~\cite{ASSA1}. $H^\pm\rightarrow\tau^\pm\nu_\tau$ extends the discovery reach to high Higgs 
masses and to lower $\tan\beta$ values in the high $\tan\beta$ region as seen in 
Fig.~\ref{fig:contour}. However, in the low $\tan\beta$ region, the $\tau^\pm\nu_\tau$ channel 
offers no sensitivity for the charged Higgs discovery as the $H^\pm\rightarrow\tau^\pm\nu_\tau$ 
branching vanishes~\cite{ASSA2}.
Below the top quark mass, the charged Higgs is produced in top decays, $t\rightarrow bH^\pm$. In 
this mass range, the decay channel $H^\pm\rightarrow\tau^\pm\nu_\tau$ has been studied for 
ATLAS and the signal appears as an excess of $\tau$ leptons~\cite{CAVA}: the entire range of 
$\tan\beta$ values should be covered for $m_{H^\pm} < m_t$ as shown in Fig.~\ref{fig:contour}. The
degradation of the sensitivity in the intermediate $\tan\beta$ region is due to suppressed charged
Higgs couplings to SM fermions as explained in section~\ref{sec:thre}. 

Charged Higgs searches might be used to probe the decoupling regime of MSSM --- hence 
distinguishing between SM and MSSM --- particularly via the $H^\pm\to\tau^\pm\nu_\tau$ channel. 
In fact, the extent of the 
parameter space that can be covered using this signature is comparable to the reach of the 
$A/H \to \tau\tau$ channel in the neutral Higgs sector, at least at large 
$\tan\beta$~\cite{DONA,TDR,ASSA2,RITVA}. Furthermore, additional improvements may still be possible
in the $H^\pm\to\tau^\pm\nu_\tau$ channel such as the possibility of exploiting the kinematics of
the spectator $b$-jet in $gg\to tbH^\pm$ and the recent calculation of a rather large k-factor for
$gb \to tH^\pm$~\cite{k-fac}.

\section{Outlook}
\label{sec:out}

In this section, we discuss additional work planned or currently being carried out in the 
charged Higgs sector to study the region below the top quark mass, the prospects for $m_{H^\pm}$ 
determination in this region, and also to cover the remaining areas of the discovery 
contour of Fig.~\ref{fig:contour}. 

\subsection{Below the top quark mass}

The $H^\pm\rightarrow\tau^\pm\nu_\tau$ channel is currently being investigated further, taking 
into account the $\tau$ polarisation effects. A direct measurement of the charged Higgs mass 
in this region is not possible because of the presence of various neutrinos in the final 
state. The possibility of measuring the charged Higgs mass from the $\tau-b$ system in 
the final state is being studied.

\subsection{Threshold Region}

\begin{figure}
\vspace*{12pt}
\begin{center}
\epsfig{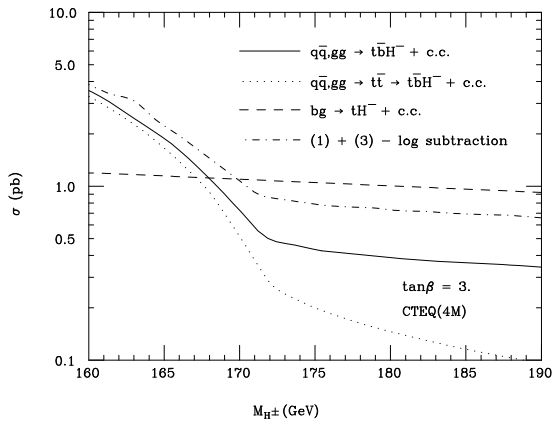}
\caption{\small Cross section for $gg,q\bar q\to t\bar b H^-$;
$gg,q\bar q \to t\bar t \to t\bar b H^-$ with finite top quark width; $bg\to tH^-$ and
the combination of the first and the last, at the LHC with $\sqrt s=14$ TeV,
as a function of $m_{H^\pm}$ for a representative value of $\tan\beta$.}
\label{fig:hthresh}
\end{center}
\end{figure}

For $m_{H^\pm}$ just below or around the top quark mass, the relevant channels are 
$H^\pm\rightarrow t^*b$ and $H^\pm\rightarrow\tau^\pm\nu_\tau$. The correct description 
of the charged Higgs production and decay mechanisms in this region of parameter space 
requires the use the production process $gg\rightarrow tbH^\pm$, shown in Fig.~\ref{fig:prod_graphs} 
which includes $gg\rightarrow t\bar{t}$ with $t\rightarrow bH^\pm$, the Higgs-strahlung mechanism 
and the relative interferences~\cite{MORE}: the narrow width approximation used by Monte Carlo 
programs such as PYTHIA~\cite{PYTHIA}, HERWIG~\cite{HERWIG} and ISAJET~\cite{ISAJET}, 
accounts for 
the charged Higgs production and decay through the factorisation approach, i.e., 
$gg,q\bar{q}\rightarrow t\bar{t}$ times $t\rightarrow bH^\pm$. However, this description does not 
account properly for the charged Higgs boson phenomenology when its mass approaches or exceeds 
that of 
the top quark as shown in Fig.~\ref{fig:hthresh}~\cite{MORE,THRE-MORE}. 
For the LHC, the situation is further complicated by the potential problem of double counting when 
adding the $2\rightarrow 3$ and the $2\rightarrow 2$ production mechanisms of 
Fig.~\ref{fig:prod_graphs}. The 5-$\sigma$ discovery contour of Fig.~\ref{fig:contour} shows a gap 
in the $m_A$ axis around $m_A=160$~GeV corresponding to the threshold region where studies 
have just 
commenced using the $gg\rightarrow tbH^\pm$ instead of the factorisation approach.
 
\subsection{Intermediate $\bm{\tan\beta}$ Region}
\label{sec:thre}

In the studies discussed thus far, a heavy SUSY spectrum has been assumed; thus charged Higgs 
decays 
into supersymmetric particles are kinematically forbidden. The lack of sensitivity in the 
intermediate $\tan\beta$ region ($\tan\beta$ $\sim$ 3-10) is due to the fact that the charged Higgs 
coupling to SM fermions is proportional to:
\begin{equation}
\label{eq:H+couplings}
H^+(m_t\cot\beta\bar{t}b_L+m_b\tan\beta\bar{t}b_R),
\end{equation}
the square of which goes through a minimum at $\tan\beta=\sqrt{m_t/m_b}$. The study of charged 
Higgs 
decays into SUSY particles might help cover this region. Indeed, for a heavy charged Higgs boson, 
the decays into the lightest charginos and neutralinos --- sleptons and squarks also --- would be 
possible and even become dominant, thereby reducing the branching into the SM decays 
$H^\pm\rightarrow tb$ and $H^\pm\rightarrow\tau^\pm\nu_\tau$~\cite{DJOU}. It was shown 
in~\cite{BISS} that $H^\pm\rightarrow\tilde{\chi}_1^\pm\tilde{\chi}^0_{\{2,3\}}$ could probe regions 
of the MSSM parameter space where the $H^\pm$ decays into SM particles yield no sensitivity --- 
see 
Fig.~\ref{fig:contour}. In addition to the direct $H^\pm$ production via the $2\rightarrow 3$ and 
the $2\rightarrow 2$ processes of Fig.~\ref{fig:prod_graphs}, the $H^\pm$ production rate in SUSY 
particle cascade decays can be significant and sensitive to the intermediate $\tan\beta$ 
values~\cite{DATTA}. Further studies of these exotic charged Higgs decays are in 
progress~\cite{MOOT}.

\subsection{High Mass Region}
The discovery reach could be extended to higher Higgs masses by studying the process 
$gg\rightarrow tbH^\pm$ with $H^\pm\rightarrow tb$ and tagging all the four $b$-jets in the 
spectrum~\cite{MOR-WH01} and exploiting the differences between the signal and the 
$gg\rightarrow t\bar{t} b\bar{b}$ background in the kinematics of $b$-quark jets: in the background, 
the $b$-quark pair produced together with $t\bar{t}$ are rather soft, collinear with low invariant 
mass. On the contrary, in the signal, at least one of the associated $b$-jets is expected to be 
energetic for $m_{H^\pm}$ much larger than $m_t$~\cite{MOR-WH01}. However, tagging four $b$-jets may 
cause a 
significant reduction in the signal rate as the additional $b$-quark in $gg\rightarrow tbH^\pm$ has a 
low transverse momentum and may escape detection. Realistic studies of this channel including 
detector effects are in progress. 

\section{Conclusions}
 
In the simplest extension of the SM Higgs sector, five Higgs bosons are predicted, three neutral and a
charged pair. The charged Higgs boson does not have a SM counterpart, thus its discovery would 
constitute an irrefutable evidence of physics beyond the Standard Model. 

In this paper, we have investigated the feasibility of the charged Higgs detection at the LHC with 
the ATLAS detector. Below the top quark mass, $H^\pm$ could be produced in the decay of the 
top quark, $t\rightarrow bH^\pm$, and the decay $H^\pm\rightarrow\tau^\pm\nu_\tau$ was 
previously 
studied for ATLAS. The signal appears as an excess of $\tau$ leptons and almost the entire range of 
$\tan\beta$ is covered. In the LEP allowed regions of MSSM parameter space, the 
$H^\pm\rightarrow W^\pm h^0$ channel presents no significant discovery potential for the charged 
Higgs. In NMSSM, LEP constraints are no longer valid and the signal viability is extended to a 
bigger area of parameter space. 

In the $H^\pm\rightarrow tb$ channel, upwards of 5-$\sigma$ discovery can be achieved above the 
top quark mass in the low and high $\tan\beta$ regions up to $\sim$400~GeV. 
The $H^\pm\rightarrow\tau^\pm\nu_\tau$ channel extends the discovery reach to higher Higgs masses and to 
lower 
$\tan\beta$ values but the sensitivity is limited to the high $\tan\beta$ region. In models with 
singlet neutrinos in large extra dimensions, the process $H^-\rightarrow\tau^-_L\psi+c.c.$ --- which 
is completely suppressed in the 2HDM --- can have a significant branching ratio, and its detection 
together with the measurement of the $\tau$ polarisation asymmetry would provide a distinctive 
evidence for these  models.

The charged Higgs mass can be determined in $H^\pm\rightarrow tb$ and 
$H^\pm\rightarrow\tau^\pm\nu_\tau$ where the precisions range from 0.5\% at $\sim$ 200~GeV to 
1.8\% 
at $\sim$ 500~GeV for an integrated luminosity of 300~fb$^{-1}$. In either channel, the main 
uncertainties come from statistical errors in the invariant mass ($H^\pm\rightarrow tb$) or the 
transverse mass ($H^\pm\rightarrow\tau^\pm\nu_\tau$) distributions. By measuring the rate of 
$H^\pm\rightarrow\tau^\pm\nu_\tau$, $\tan\beta$ can be determined with precisions ranging from 
7.4\% 
at $\tan\beta=20$ to 5.4\% at $\tan\beta=50$ for an integrated luminosity of 300~fb$^{-1}$ and 
assuming a 10\% uncertainty on the luminosity. 

Further charged Higgs studies are planned or are currently being carried out in order to cover the 
remaining regions of the parameter space. These include: firstly, the threshold region where the 
$2\rightarrow 3$ process is used to correctly account for the $H^\pm$ production and decay 
phenomenology; secondly, the intermediate $\tan\beta$ region which is sensitive to charged 
Higgs decays to SUSY particles; and finally, the high mass region which could be probed with 
$gg\rightarrow tbH^\pm$ and $H^\pm\rightarrow tb$ by tagging all the four $b$-jets in the spectrum. 
In addition, the region below the top quark mass is being investigated using 
$H^\pm\rightarrow\tau^\pm\nu_\tau$ taking into account the $\tau$-polarisation effects, and the 
prospects for $m_{H^\pm}$ determination in this region is also being studied.  

\begin{acknowledgments} 

We express gratitude to E.~Richter-W\c{a}s, K.~Jacobs and the ATLAS 
Higgs working group for fruitful discussions and constructive criticisms. This work was mainly 
performed within the ATLAS collaboration and we thank the collaboration members for helpful 
discussions. Parts of this work were carried out at the Les Houches Workshop: ``Physics at TeV 
Colliders'' 1999 and 2001. We thank the organisers for the invitation and for their effort. 
K.~A.~Assamagan thanks  K.~Agashe for helpful correspondence, and S.~Moretti and R.~Kinnunen 
for useful discussions. K.~A.~Assamagan's contribution was partially supported by grants from 
the US National Science Foundation (grant numbers 9722827 and PHY-0072686) at Hampton University 
prior to his appointment at Brookhaven National Laboratory.  
\end{acknowledgments}


\begin{thebibliography}{99} 

\bibitem{2HDM}
J.F.~Gunion, H.E.~Haber, G.L.~Kane, S.~Dawson, The Higgs Hunter's Guide (Addison-Wesley, 
Reading, MA 1990), 
Erratum: arXiv:hep-ph/9302272;
H.~P.~Nilles,
Phys.\ Rept.\  {\bf 110}, 1 (1984);
H.~E.~Haber and G.~L.~Kane,
Phys.\ Rept.\  {\bf 117}, 75 (1985).

\bibitem{CONST}
S.~Heinemeyer, W.~Hollik and G.~Weiglein,
Eur.\ Phys.\ J.\ C {\bf 9}, 343 (1999)
[arXiv:hep-ph/9812472].

\bibitem{LEP-SM}
The ALEPH, DELPHI, L3 and OPAL Collaborations and the LEP Higgs Working Group, 
CERN-EP/2001-055, arXiv:hep-ex/0107030.

\bibitem{GFIT}
The LEP Electroweak Working Group, A Combination of Preliminary Electroweak
Measurements and  Constraints in the Standard Model, CERN-EP/2001- in
preparation, presented by D.~Charlton, EPS  HEP 
2001, Budapest, Hungary, July 12-18, 2001.

\bibitem{h_TeV}
L.~Moneta, talk given at the `36th Rencontres de Moriond on QCD and Hadronic
Interactions',  Les Arcs, France, 17-24 March 2001, arXiv:hep-ex/0106050.

\bibitem{OPAL} G.~Abbiendi, OPAL Collaboration, Eur. \ Phys. \ J. \ C {\bf 18}, 425, (2001) 
[arXiv:hep-ex/0007040].


\bibitem{Hp_RAD}
A.~Brignole, J.~Ellis, G.~Ridolfi, F.~Zwirner, Phys. Lett. B \textbf{271}, 123 (1991); 
A.~Brignole, Phys. Lett. B \textbf{277}, 313 (1992); 
M.~A. D\'{i}az, H.~E.~Haber, Phys Rev. D \textbf{45}, 4246 (1992).

\bibitem{hp_LEP} ALEPH, DELPHI, L3 and OPAL Collaborations, CERN-EP/2000-055; ALEPH 
Collaboration, CERN-EP/2000-086, Phys. Lett. B487 (2000) 253; 
DELPHI Collaboration, 
CERN-EP/2001-062, M.~Ellert et al., Nucl. Phys. B -- Proceeding supplements 98 (2001), 336; 
L3 Collaboration, CERN-EP/2000-118, Phys. Lett. B496 (2000) 34; OPAL Collaboration, 
CERN-EP/98-173.

\bibitem{hp_TEV}
P.~Gutierrez, FERMILAB-Conf-00-294-E; 
V.~M.~Abazov {\it et al.}  [D{\O} Collaboration],
arXiv:hep-ex/0102039.

\bibitem{CLEO}
M.~S.~Alam {\it et al.}  [CLEO Collaboration],
Phys.\ Rev.\ Lett.\  {\bf 74}, 2885 (1995);
R.Briere, in Proceedings of ICHEP98, Vancouver Canada, 1998, CLEO-CONF-98-17; ibid. and in 
talk by J.~Alexander; R.~Barate {\it et al.}  [ALEPH Collaboration],
Phys.\ Lett.\ B {\bf 429}, 169 (1998).

\bibitem{Gambino:2001ew}
P.~Gambino and M.~Misiak,
Nucl.\ Phys.\ B {\bf 611} 338 (2001)

\bibitem{BORZ}
F.~M.~Borzumati and C.~Greub,
Phys.\ Rev.\ D {\bf 58}, 074004 (1998)
[arXiv:hep-ph/9802391];
Phys.\ Rev.\ D {\bf 59}, 057501 (1999)
[arXiv:hep-ph/9809438].

\bibitem{COAR}
J.~A.~Coarasa, J.~Guasch, J.~Sola and W.~Hollik,
Phys.\ Lett.\ B {\bf 442}, 326 (1998)
[arXiv:hep-ph/9808278].

\bibitem{Hp-PROD}
J.~F.~Gunion, H.~E.~Haber, F.~E.~Paige, W.~K.~Tung and S.~S.~Willenbrock,
Nucl.\ Phys.\ B {\bf 294}, 621 (1987);
J.~L.~Diaz-Cruz and O.~A.~Sampayo,
Phys.\ Rev.\ D {\bf 50}, 6820 (1994).

\bibitem{Hp-PROD1}
A.~Krause, T.~Plehn, M.~Spira and P.~M.~Zerwas,
Nucl.\ Phys.\ B {\bf 519}, 85 (1998)
[arXiv:hep-ph/9707430];
Y.~Jiang, W.~g.~Ma, L.~Han, M.~Han and Z.~h.~Yu,
J.\ Phys.\ G {\bf 24}, 83 (1998)
[arXiv:hep-ph/9708421];
A.~A.~Barrientos Bendezu and B.~A.~Kniehl,
Nucl.\ Phys.\ B {\bf 568}, 305 (2000)
[arXiv:hep-ph/9908385];
O.~Brein and W.~Hollik,
Eur.\ Phys.\ J.\ C {\bf 13}, 175 (2000)
[arXiv:hep-ph/9908529].

\bibitem{Hp-PROD2}
D.~A.~Dicus, J.~L.~Hewett, C.~Kao and T.~G.~Rizzo,
Phys.\ Rev.\ D {\bf 40}, 787 (1989);
A.~A.~Barrientos Bendezu and B.~A.~Kniehl,
Phys.\ Rev.\ D {\bf 59}, 015009 (1999)
[arXiv:hep-ph/9807480];
Phys.\ Rev.\ D {\bf 61}, 097701 (2000)
[arXiv:hep-ph/9909502];
Phys.\ Rev.\ D {\bf 63}, 015009 (2001)
[arXiv:hep-ph/0007336];
O.~Brein, W.~Hollik and S.~Kanemura,
Phys.\ Rev.\ D {\bf 63}, 095001 (2001)
[arXiv:hep-ph/0008308].

\bibitem{WH} S.~Moretti and K.~Odagiri, Phys. Rev. D {\bf 59}, 055008, (1999).

\bibitem{subtraction} 
F.~Borzumati, J.~L.~Kneur and N.~Polonsky,
Phys.\ Rev.\ D {\bf 60}, 115011 (1999)
[arXiv:hep-ph/9905443].

\bibitem{3b}
S.~Moretti and D.~P.~Roy,
Phys.\ Lett.\ B {\bf 470}, 209 (1999)
[arXiv:hep-ph/9909435].

\bibitem{reviewLHC} 
A.~Djouadi {\it et al.},
``The Higgs working group: Summary report,''
Proceedings of the Workshop `Physics at TeV Colliders',
Les Houches, France, 8-18 June 1999,
arXiv:hep-ph/0002258;
D.~Denegri {\it et al.},
arXiv:hep-ph/0112045.

\bibitem{Jaume} 
A.~Belyaev, D.~Garcia, J.~Guasch and J.~Sola,
Phys.\ Rev.\ D {\bf 65}, 031701 (2002)
[arXiv:hep-ph/0105053].

\bibitem{HDECAY}  
A.~Djouadi, J.~Kalinowski and M.~Spira,
Comput.\ Phys.\ Commun.\  {\bf 108}, 56 (1998)
[arXiv:hep-ph/9704448].

\bibitem{ASSA1} 
K.~A.~Assamagan,
Acta Phys.\ Polon.\ B {\bf 31}, 863 (2000); ATLAS Internal Note
 ATL-PHYS-99-013.

\bibitem{ASSA2}  
K.A.~Assamagan and Y.~Coadou, Acta Phys.\ Polon.\ B {\bf 33}, 707 (2002); 
ATLAS Internal Note ATL-PHYS-2000-031 (2000).

\bibitem{ASSA3} 
K.A.~Assamagan, 
Acta Phys.\ Polon.\ B {\bf 31}, 881 (2000); ATLAS Internal
Note ATL-PHYS-99-025.  

\bibitem{ASSA4} 
K.A.~Assamagan and Y.~Coadou, Acta Phys.\ Polon.\ B {\bf 33}, 1347 (2002); 
ATLAS Internal Note ATL-PHYS-2001-017 (2001). 

\bibitem{ASSA5} 
K.~A.~Assamagan and A.~Deandrea, ATLAS Internal Note ATL-PHYS-2001-019,
Phys. \ Rev. \ D {\bf 65}, 076006 (2002) [arXiv:hep-ph/0111256].

\bibitem{PYTHIA}  
T.~Sjostrand,
Comput.\ Phys.\ Commun.\  {\bf 82}, 74 (1994);
T.~Sjostrand, P.~Eden, C.~Friberg, L.~Lonnblad, G.~Miu, S.~Mrenna and E.~Norrbin,
Comput.\ Phys.\ Commun.\  {\bf 135}, 238 (2001)
[arXiv:hep-ph/0010017];
T.~Sjostrand, L.~Lonnblad and S.~Mrenna,
arXiv:hep-ph/0108264.

\bibitem{ATLFAST}  
E.~Richter-W\c{a}s, 
D.~Froidevaux and L.~Poggioli, ATLAS Internal Note, ATL-PHYS-98-131, (1998). 

\bibitem{CTEQ}  
H.~L.~Lai {\it et al.},
Phys.\ Rev.\ D {\bf 55}, 1280 (1997)
[arXiv:hep-ph/9606399].

\bibitem{FeynHiggs} 
S.~Heinemeyer, W.~Hollik and G.~Weiglein, 
arXiv:hep-ph/0002213. 

\bibitem{TP} 
ATLAS Collaboration, Technical Proposal for a General Purpose pp 
Experiment at the Large Hadron Collider at CERN, CERN/LHCC/94-43, LHCC/P2, 15
December 1994.

\bibitem{TDR} 
ATLAS Collaboration, 'ATLAS Detector and Physics Performance
Technical Design  Report',CERN/LHCC/99-15, 25 May (1999) 285.

\bibitem{BOSM} 
M.~Bosman, Talk given at the IX International Conference on Calorimetry in 
Particle Physics, October 9-14, 2000, Annecy, France, ATL-CONF-2002-001 (2002)

\bibitem{barger}
V.~D.~Barger, R.~J.~Phillips and D.~P.~Roy,
Phys.\ Lett.\ B {\bf 324}, 236 (1994)
[arXiv:hep-ph/9311372].

\bibitem{oda}
K.~Odagiri, RAL-TR-1999-012; hep-ph/9901432.

\bibitem{POL1} 
D.~P.~Roy,
Phys.\ Lett.\ B {\bf 459}, 607 (1999)
[arXiv:hep-ph/9905542].

\bibitem{POL2} 
B.~K.~Bullock, K.~Hagiwara and A.~D.~Martin,
Nucl.\ Phys.\ B {\bf 395}, 499 (1993).

\bibitem{POL3} 
S.~Raychaudhuri and D.~P.~Roy,
Phys.\ Rev.\ D {\bf 52}, 1556 (1995)
[arXiv:hep-ph/9503251];
Phys.\ Rev.\ D {\bf 53}, 4902 (1996)
[arXiv:hep-ph/9507388].

\bibitem{TAUOLA} 
S.~Jadach, Z.~W\c{a}s, R.~Decker and J.~H.~K\"{u}hn, 
Comput.\ Phys.\ Commun.\  {\bf 76}, 361 (1993); 
M.~Jezabek, Z.~W\c{a}s, S.~Jadach and J.~H.~K\"{u}hn, 
Comput.\ Phys.\ Commun.\  {\bf 70}, 69 (1992); 
S.~Jadach, J.~H.~K\"{u}hn and Z.~W\c{a}s, 
Comput.\ Phys.\ Commun.\  {\bf 64}, 275 (1990). 

\bibitem{MOR-WH01} 
D.~J.~Miller, S.~Moretti, D.~P.~Roy and W.~J.~Stirling,
Phys.\ Rev.\ D {\bf 61}, 055011 (2000)
[arXiv:hep-ph/9906230]. 

\bibitem{MOR-WH02} 
S.~Moretti and W.~J.~Stirling,
Phys.\ Lett.\ B {\bf 347}, 291 (1995)
[Erratum-ibid.\ B {\bf 366}, 451 (1996)]
[arXiv:hep-ph/9412209]; 
A. Djouadi, J. Kalinowski and P.M. Zerwas, Z. Phys. {C70} (1996) 435; 
E. Ma, D.P. Roy and J. Wudka, Phys. Rev. Lett. {80} (1998) 1162.

\bibitem{MOR-WH03} 
M.~Drees, M.~Guchait and D.~P.~Roy,
Phys.\ Lett.\ B {\bf 471}, 39 (1999)
[arXiv:hep-ph/9909266]
and the references therein.

\bibitem{Wh0} S.~Moretti, Phys. Lett. B {\bf 481}, 49, (2000) [arXiv:hep-ph/0003178].

\bibitem{Akero1} A.~G.~Akeroyd and S.~Baek, Phys.\ Lett. \ B {\bf 525}, 315 (2002)
[arXiv:hep-ph/0105228].

\bibitem{Akero2} A.~G.~Akeroyd, A.~Arhrib and E.~Naimi, Eur. \ Phys. \ J \ C {\bf 20}, 51 (2001) 
[arXiv:hep-ph/0002288].

\bibitem{RITVA} 
R.~Kinnunen, ''The $H^\pm\rightarrow\tau^\pm\nu_\tau$ mode in CMS'' and 
''Signatures of Heavy Charged Higgs Bosons at the LHC'', Les Houches Workshop
(1999), hep-ph/0002258; D.~Denegri  et al., 
CMS Internal Note CMS-NOTE-2001-032, hep-ph/0112045.

\bibitem{HOHL} 
M.~Hohlfeld, ATLAS Internal Note ATL-PHYS-2001-004 (2001).

\bibitem{GUNI} 
J.~F.~Gunion, L.~Poggioli, R.~Van Kooten, C.~Kao and P.~Rowson,
arXiv:hep-ph/9703330.

\bibitem{KK} 
T.~Kaluza, 
Sitzungsber.\ Preuss.\ Akad.\ Wiss.\ Berlin (Math.\ Phys.\ ) {\bf K1}, 966 
(1921);  
for a translation of the original paper see T.~Muta, HUPD-8401 
{\it  in O'Raifeartaigh, L.: The dawning of gauge theory* 53-58};
O.~Klein, 
Z.\ Phys.\  {\bf 37}, 895 (1926) 
[Surveys High Energ.\ Phys.\  {\bf 5}, 241 (1926)]. 
 
\bibitem{1} 
N.~Arkani-Hamed, S.~Dimopoulos and G.~R.~Dvali, 
Phys.\ Lett.\ B {\bf 429}, 263 (1998) 
[arXiv:hep-ph/9803315]; 
Phys.\ Rev.\ D {\bf 59}, 086004 (1999) 
[arXiv:hep-ph/9807344]; 
I.~Antoniadis, N.~Arkani-Hamed, S.~Dimopoulos and G.~R.~Dvali, 
Phys.\ Lett.\ B {\bf 436}, 257 (1998) 
[arXiv:hep-ph/9804398]; 
G.~F.~Giudice, R.~Rattazzi and J.~D.~Wells, 
Nucl.\ Phys.\ B {\bf 544}, 3 (1999) 
[arXiv:hep-ph/9811291]; 
E.~A.~Mirabelli, M.~Perelstein and M.~E.~Peskin, 
Phys.\ Rev.\ Lett.\  {\bf 82}, 2236 (1999) 
[arXiv:hep-ph/9811337]; 
T.~Han, J.~D.~Lykken and R.~J.~Zhang, 
Phys.\ Rev.\ D {\bf 59}, 105006 (1999) 
[arXiv:hep-ph/9811350]. 
 
\bibitem{2} 
N.~Arkani-Hamed, S.~Dimopoulos, G.~R.~Dvali and J.~March-Russell, 
arXiv:hep-ph/9811448. 
K.~R.~Dienes, E.~Dudas and T.~Gherghetta, 
Nucl.\ Phys.\ B {\bf 557}, 25 (1999) 
[arXiv:hep-ph/9811428]. 

\bibitem{rub}
K.~Akama,
Lect.\ Notes Phys.\  {\bf 176}, 267 (1982)
[arXiv:hep-th/0001113];
V.~A.~Rubakov and M.~E.~Shaposhnikov,
Phys.\ Lett.\ B {\bf 125} (1983) 136;
for a review see V.~A.~Rubakov,
``Large and infinite extra dimensions: An Introduction,''
arXiv:hep-ph/0104152.

\bibitem{pol}
J.~Polchinski,
arXiv:hep-th/9611050.
C.~P.~Bachas,
arXiv:hep-th/9806199.


\bibitem{3}  
S.~P.~Martin and J.~D.~Wells, 
Phys.\ Rev.\ D {\bf 60}, 035006 (1999) 
[arXiv:hep-ph/9903259]. 
 
\bibitem{4} 
Y.~Fukuda {\it et al.}  [SuperKamiokande Collaboration], 
Phys.\ Rev.\ Lett.\  {\bf 82}, 2644 (1999) 
[arXiv:hep-ex/9812014]; 
{\bf 85}, 3999 (2000) 
[arXiv:hep-ex/0009001]; 
{\bf 86}, 5656 (2001) 
[arXiv:hep-ex/0103033]; 
A.~Ioannisian and A.~Pilaftsis, 
Phys.\ Rev.\ D {\bf 62}, 066001 (2000) 
[arXiv:hep-ph/9907522]. 
R.~N.~Mohapatra and A.~Perez-Lorenzana, 
Nucl.\ Phys.\ B {\bf 576}, 466 (2000) 
[arXiv:hep-ph/9910474]; 
A.~S.~Dighe and A.~S.~Joshipura, 
Phys.\ Rev.\ D {\bf 64}, 073012 (2001) 
[arXiv:hep-ph/0105288];
A.~De Gouvea, G.~F.~Giudice, A.~Strumia and K.~Tobe,
arXiv:hep-ph/0107156.

\bibitem{agashe} 
K.~Agashe, N.~G.~Deshpande and G.~H.~Wu, 
Phys.\ Lett.\ B {\bf 489}, 367 (2000) 
[arXiv:hep-ph/0006122]. 

\bibitem{CAVA} 
D.~Cavalli et al., ATLAS Internal Note ATL-PHYS-94-053 (1994).

\bibitem{DONA} D.~Cavalli and S.~Resconi, ATLAS Internal Note ATL-PHYS-2000-005 (2000).

\bibitem{k-fac} S.-H.~Zhu, arXiv:hep-ph/0112109.

\bibitem{MORE} 
M.~Guchait and S.~Moretti,
JHEP {\bf 0201}, 001 (2002)
[arXiv:hep-ph/0110020].

\bibitem{HERWIG} 
G.~Marchesini, B.~R.~Webber, G.~Abbiendi, I.~G.~Knowles, M.~H.~Seymour and L.~Stanco,
Comput.\ Phys.\ Commun.\  {\bf 67}, 465 (1992).
G.~Corcella {\it et al.},
arXiv:hep-ph/9912396.
G.~Corcella {\it et al.},
arXiv:hep-ph/0107071, arXiv:hep-ph/0201201.

\bibitem{ISAJET} 
H.~Baer, F.~E.~Paige, S.~D.~Protopopescu and X.~Tata,
arXiv:hep-ph/0001086.

\bibitem{THRE-MORE} 
S.~Moretti, private communication.

\bibitem{DJOU} 
F.~Borzumati and A.~Djouadi, hep-ph/9806301; P.~M.~Zerwas et al., ECFA-DESY 
Workshop, hep-ph/9605437.

\bibitem{BISS} 
M.~Bisset, M.~Guchait and S.~Moretti,
Eur.\ Phys.\ J.\ C {\bf 19}, 143 (2001)
[arXiv:hep-ph/0010253].

\bibitem{DATTA} 
A.~Datta, A.~Djouadi, M.~Guchait and Y.~Mambrini,
Phys.\ Rev.\ D {\bf 65}, 015007 (2002)
[arXiv:hep-ph/0107271].

\bibitem{MOOT} 
M. Bisset, F. Moortgat and S. Moretti, in preparation.


\end{thebibliography}
\end{document}